\documentclass[lettersize,journal]{IEEEtran}
\usepackage{amsmath,amsfonts}
\usepackage{algorithmic}
\usepackage{algorithm}
\usepackage{xcolor}
\usepackage{array}
\usepackage{subcaption}
\usepackage{textcomp}
\usepackage{stfloats}
\usepackage{url}
\usepackage{verbatim}
\usepackage{graphicx}
\def\BibTeX{{\rm B\kern-.05em{\sc i\kern-.025em b}\kern-.08emT\kern-.1667em\lower.7ex\hbox{E}\kern-.125emX}}
\usepackage{balance}
\usepackage{cite}
\usepackage{xurl}

\begin{document}

\title{Constructing Per-Shot Bitrate Ladders using Visual Information Fidelity}

\author{
    Krishna Srikar Durbha and Alan C. Bovik,~\IEEEmembership{Life Fellow,~IEEE}
    \thanks{
        This research was sponsored by grant number 2019844 from the National Science Foundation AI Institute for Foundations of Machine Learning (IFML).
    }
    \thanks{
        Krishna Srikar Durbha is with the Laboratory for Image and Video Engineering (LIVE), The University of Texas at Austin, Austin, TX 78712, USA (Email: krishna.durbha@utexas.edu).
    }
    \thanks{
        Alan C. Bovik is with the Colorado's Laboratory for Image and Video Engineering (LIVE), University of Colorado Boulder, Boulder, CO 80309 USA (e-mail: Alan.Bovik@colorado.edu).
    }
    \thanks{
        % All databases accessed during training and evaluation were performed at The University of Texas at Austin by university faculty, staff, and students.
        The implementations used in this work have been made available at https://github.com/krishnasrikard/Constructing-Per-Shot-Bitrate-Ladders-using-Visual-Information-Fidelity.
    }
}

\markboth{Journal of \LaTeX\ Class Files, ~Vol.~xx, No.~x, Month~20xx}{}

\maketitle

\begin{abstract}
    Video service providers need their delivery systems to be able to adapt to network conditions, user preferences, display settings, and other factors. HTTP Adaptive Streaming (HAS) offers dynamic switching between different video representations to simultaneously enhance bandwidth consumption and users' streaming experiences. Per-shot encoding, pioneered by Netflix, optimizes the encoding parameters on each scene or shot. The Dynamic Optimizer (DO) uses the Video Multi-Method Assessment Fusion (VMAF) perceptual video quality prediction engine to deliver high-quality videos at reduced bitrates. Here we develop a perceptually optimized method of constructing optimal per-shot bitrate and quality ladders, using an ensemble of low-level features and Visual Information Fidelity (VIF) features. During inference, our method predicts the bitrate or quality ladder of a source video without any compression or quality estimation. We compare the performance of our model against other content-adaptive bitrate ladder prediction methods, a fixed bitrate ladder, and reference bitrate ladders constructed via exhaustive encoding using Bj{\o}ntegaard-delta (BD) metrics. Our proposed method shows excellent gains in bitrate and quality against the fixed bitrate ladder and only small losses against the reference bitrate ladder, while providing significant computational advantages.
\end{abstract}

\begin{IEEEkeywords}
    Adaptive Streaming, Bitrate Ladder Construction, Video Processing, Gaussian Mixture Models
\end{IEEEkeywords}

\section{Introduction}
\label{sec:introduction}
A recent report on video streaming \cite{Video-Streaming-Report} stated that as of 2024, video constitutes about 74\% of mobile traffic, and is predicted to increase more than 80\% by 2029. While there is a noticeable increase in user-generated content (UGC) uploaded to social media platforms, the majority of video traffic can be attributed to Video-on-Demand (VoD) services offered by Netflix, Meta, YouTube, Prime Video, and others. VoD service providers deliver videos to users that are scaled and/or compressed based on their display settings, network conditions, device capabilities, available bandwidth, and buffer state. Video service providers invest substantial resources towards optimizing video compression and delivery pipelines both to decrease video transmission costs and to enhance end-user satisfaction.

In recent years, \textit{HTTP Adaptive Streaming} has emerged as an effective and popular standard for video content delivery, and \textit{HTTP Live Streaming} (HLS) bitrate ladders \cite{Fixed-Bitrate-Ladder} have become widely adopted go-to ways of adaptive bitrate streaming of all kinds of video content. HLS \cite{Fixed-Bitrate-Ladder}, also known as `one-size-fits-all' is designed to allow adaptation to various video characteristics and network conditions. It allows the selection of different sets of video encoding parameters, based on network conditions and user preferences, to ensure a high \textit{Quality of Experience} (QoE) when displayed. However, a disadvantage of HLS \cite{Fixed-Bitrate-Ladder} is that the encoding settings remain the same, independent of changes in the video content, thereby not guaranteeing an optimal bitrate ladder for a given video.

Recently introduced content-optimized per-shot encoding techniques \cite{Per-Title-Encoding, Shot-Encoding, Dynamic-Optimizer} provide bitrate savings and better QoE than existing fixed bitrate ladders (HLS). Per-shot encoding techniques achieve optimal encoding by constructing a convex hull on each shot of the video. Each video title to be streamed is partitioned into shorter scenes or shots of relatively shorter durations which are encoded independently of one another. Generally, shots consist of frames that are relatively homogenous in content, hence suitable for encoding using locally fixed parameters. Each shot is encoded multiple times using a variety of bitrate-optimized encoding parameters, usually under a perceptual quality criterion, to construct a convex hull. The convex hull is where an encoding point reaches Pareto efficiency: it consists of optimal bitrate-resolutions pairs that display the highest perceptual quality at each of a set of typical bitrates. Given the extensive space of encoding settings, including spatial resolutions, bitrates, quantization parameters (QPs), and constant rate factors (CRFs), each shot is compressed under a quality criterion such as SSIM \cite{SSIM} multiple times when constructing the convex hull, which requires a significant amount of resources and time. For example, assuming a single shot, and given a specific video codec, a set of $R$ resolutions, and $B$ bitrates, constructing a Pareto-front requires performing compression and quality estimation $R \times B$ times. To achieve optimal performance on transmission of a single diverse video containing multiple shots/scenes, the convex hull construction needs to be repeated many times. This, in the context of per-shot encoding, which necessitates the construction of a convex hull on each individual shot, the terms `shot' and `video' along with `convex hull' and `Pareto front' will be used interchangeably throughout.

\begin{figure*}
    \centering
    \begin{subfigure}[b]{0.49\textwidth}
        \centering
        \includegraphics[width=\textwidth]{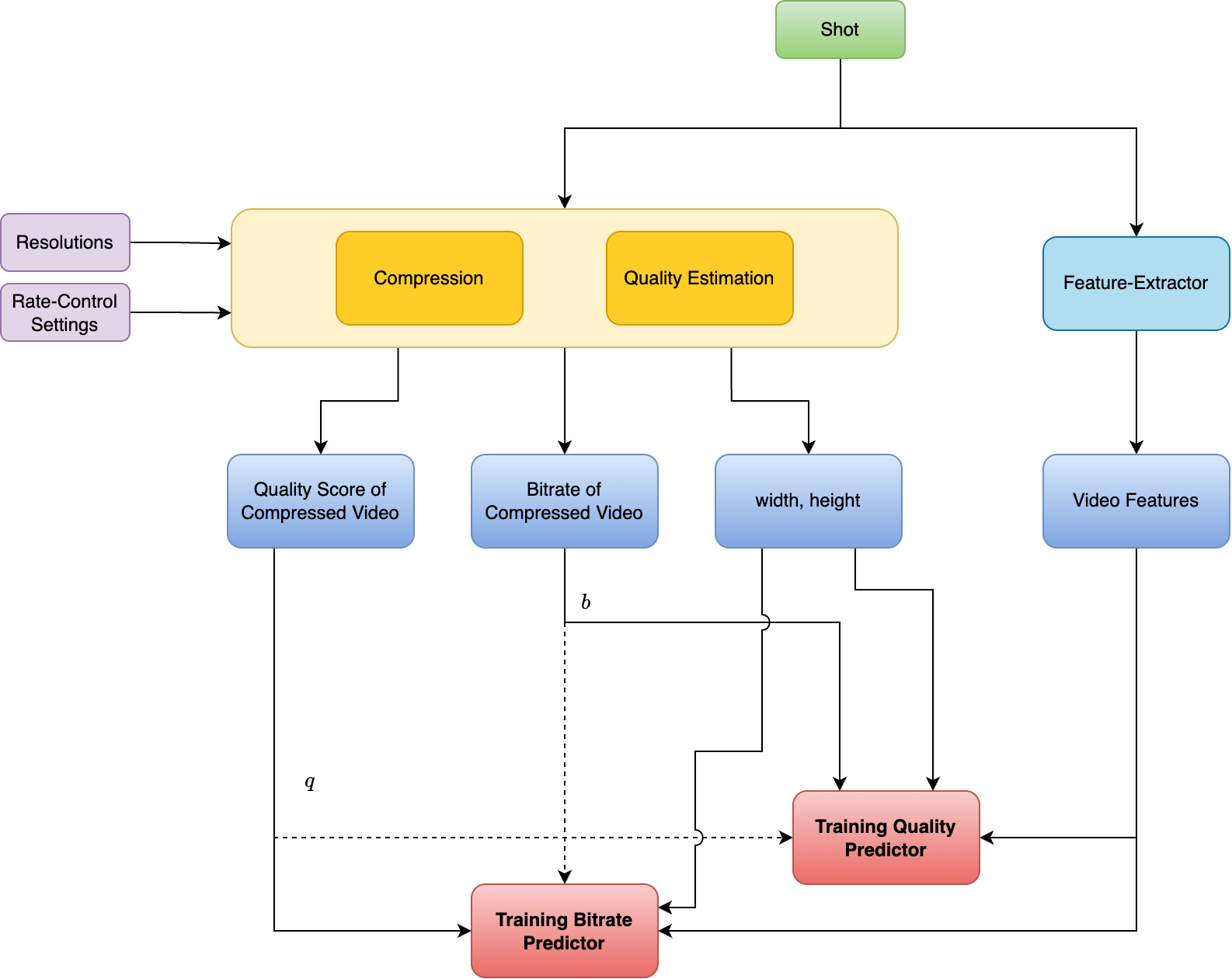}
        \caption{Proposed framework for training}
        \label{fig:Proposed-Framework-Training}
    \end{subfigure}
    \hfill
    \begin{subfigure}[b]{0.49\textwidth}
        \centering
        \includegraphics[width=\textwidth]{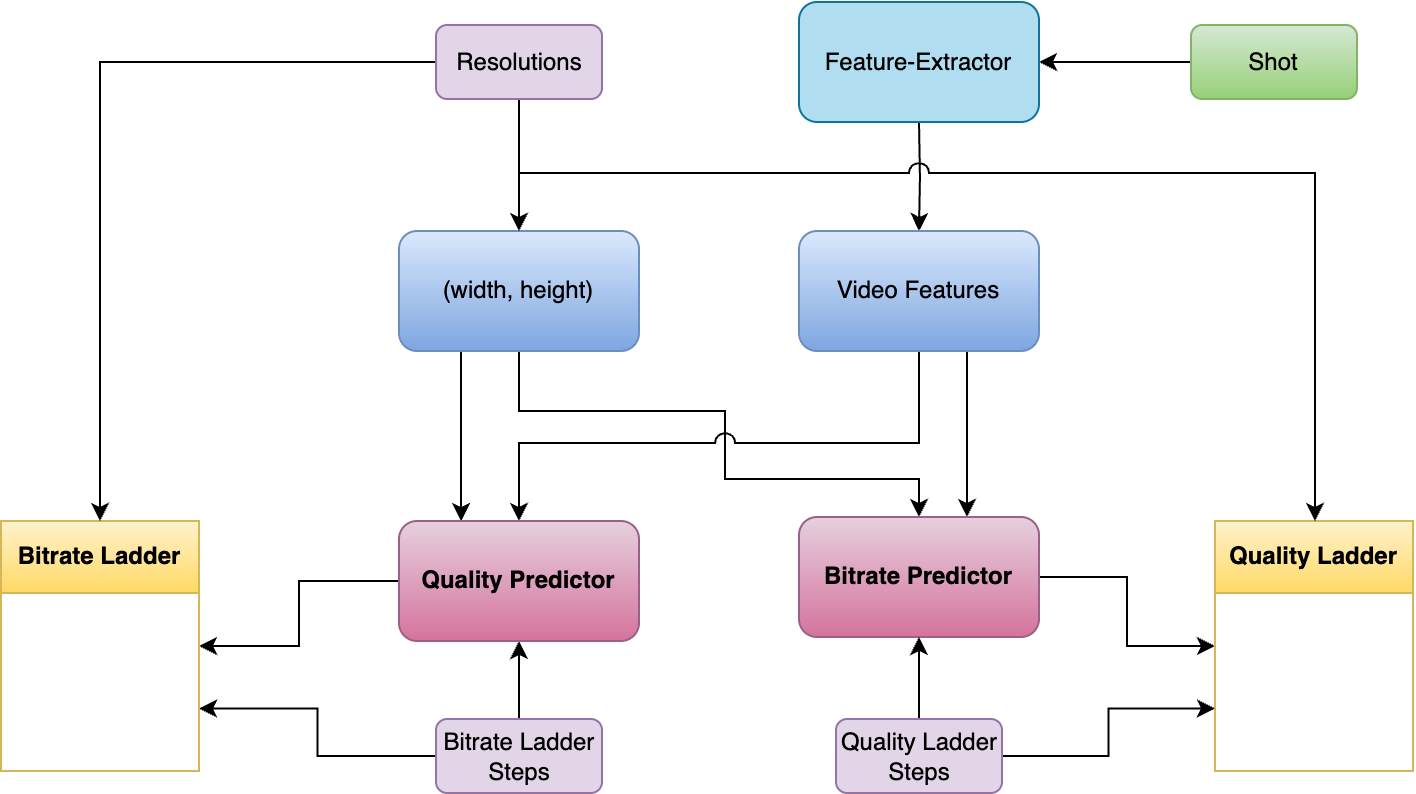}
        \caption{Proposed framework for inference}
        \label{fig:Proposed-Framework-Testing}
    \end{subfigure}
    \caption{Proposed framework for training and inference.}
    \label{fig:Proposed-Framework}
\end{figure*}

Considering the above, we present content-gnostic techniques for predicting optimal bitrate and quality ladders for various adaptive streaming services. Our proposed methods extract features from uncompressed videos at their original resolution without any compression or quality estimation, and use machine learning models to predict the quality or bitrate of a compressed video at a given resolution and bitrate or quality, respectively. With these predicted quality or bitrate values, we construct a suitable bitrate or quality ladder that yields performance comparable to a reference bitrate ladder for a given video. The reference bitrate ladder of a video is sampled from the Pareto front, which is constructed by exhaustive encoding of the video at different resolutions and rate control settings. Figures \ref{fig:Proposed-Framework-Training} and \ref{fig:Proposed-Framework-Testing} illustrate the training and inference flows of our methods. We will elaborate on these procedures in the subsequent sections. Our proposed methods offer a distinct advantage in terms of reduced complexity, as compared to the compute-intensive process involved in constructing a reference bitrate ladder.

\subsection{Contributions}
In preliminary work \cite{Bitrate-Ladder-Construction-using-Visual-Information-Fidelity} presented in the conference paper, we experimented with features drawn from Visual Information Fidelity (VIF) \cite{VIF} to predict the quality of videos compressed by libx265, given resolution and bitrate. We utilized multiple VIF feature sets extracted from different scales and subbands of a video to predict per-shot bitrate ladders. We extracted bitrate and VMAF scores from videos, compressed using \textbf{libx265} with \textbf{medium} preset, across eight different spatial resolutions and 33 different CRFs, to train an Extra-Trees \cite{Extra-Trees} regressor that predicts the VMAF scores of compressed videos based on VIF features, bitrate, width, and height. We later used the regressor to create a per-shot bitrate ladder for a given video. We have since greatly expanded this preliminary work, making the following new contributions:
\begin{itemize}
    \item We study the effectiveness of our previous quality prediction models under specific quality constraints.
    \item Similar to quality prediction models, we develop bitrate prediction models using the same VIF features to construct per-shot quality ladders.
    \item We study the performance of an ensemble of low-level features and VIF feature sets on predicting the quality and bitrate of compressed videos, and subsequently, on predicting per-shot bitrate and quality ladders, respectively.
    \item Furthermore, we compare the effectiveness of our techniques against popular content-adaptive bitrate ladder construction methods. We also design counterparts of these existing models, configured to instead predict quality ladders. These models are also included in our comparative study.
    \item Finally, we report the performance of the compared models against standard methods, including fixed bitrate ladders and reference bitrate ladders constructed by exhaustive encoding.
\end{itemize}

\subsection{Paper Organization}
The rest of the paper is organized as follows. Section \ref{sec:related_works} discusses previous work on the construction of per-shot bitrate ladders. Section \ref{sec:method} explains the computation of quality-aware VIF feature sets and content-aware low-level features and how they are used during training and inferencing in our model. Section \ref{sec:experiments-and-results} discusses the dataset and experimental settings we used in our simulations, the performances of quality and bitrate prediction regressors, their effectiveness in constructing per-shot bitrate/quality ladders, and experimental comparisons with prior methods. The paper concludes in Section \ref{sec:conclusion} along with a discussion of future directions of research.

\section{Related Work}
\label{sec:related_works}
As discussed in the Introduction, although a fixed bitrate ladder is designed considering various video characteristics, network conditions, resolutions, and bitrates, it is still content-independent. Despite the noteworthy performance of these `one-size-fits-all' fixed bitrate ladders across a broad spectrum of videos, optimization of the encoding settings at the shot level makes it possible to further enhance the Quality of Experience (QoE) while also achieving bitrate savings. The per-shot encoding framework introduced by Netflix \cite{Per-Title-Encoding, Dynamic-Optimizer, Shot-Encoding} performs search in encoding parameter space to determine optimal encoding settings for each shot. Over the past few years, a variety of techniques have been introduced to facilitate the construction of per-shot bitrate ladders, towards reducing reliance on exhaustive encoding methods.

For example, the authors of \cite{Predicting-Video-Rate-Distortion-Curves-using-Textural-Features}, modeled PSNR as a linear function of bitrate. They extracted features including gray-level co-occurrence matrices (GLCM) \cite{GLCM}, temporal coherence (TC) \cite{TC} measures, and normalized cross-correlation (NCC) to predict the coefficients of the linear function using a support vector regressor (SVR). The same authors extended their work in \cite{Study-of-compression-statistics-and-prediction-of-rate-distortion-curves-for-video-texture} by considering a bigger feature set and by investigating the ability of various parametric functions to fit Rate-Quality (RQ) curves. They concluded that although third-degree polynomials fit rate-quality (RQ) curves well, an exponential function was able to achieve better Bj{\o}ntegaard-delta (BD) \cite{BD} rate savings with less complexity. In \cite{Content-gnostic-Bitrate-Ladder-Prediction-for-Adaptive-Video-Streaming} and \cite{Efficient-Bitrate-Ladder-Construction-for-Content-Optimized-Adaptive-Video-Streaming}, the authors constructed optimal Rate-PSNR curves using cross-over points. At each cross-over point, switching from a lower resolution to a higher resolution occurs given increases in bitrate. The authors modeled cross-over points between two resolutions as a pair of QPs, with a single QP defined at each resolution. Low-level features like GLCM and TC are used to predict cross-over QPs. The authors in \cite{VMAF-based-Bitrate-Ladder-Estimation-for-Adaptive-Streaming} used `knee-points' to construct optimal Rate-VMAF curves. The knee points were defined as QPs where a rate-quality curve has the highest curvature. They used features like the GLCM and NCC to predict the knee QPs. Instead of modeling cross-over points as QPs, the authors in \cite{Benchmarking-Learning-based-Bitrate-Ladder-Prediction-Methods-for-Adaptive-Video-Streaming} modeled cross-over points as bitrates. They experimented with shallow machine learning models like Extra-Trees regressor \cite{Extra-Trees}, XGBoost \cite{XGBoost}, and Gaussian Processes \cite{GP} regression, using features like GLCM, TC, spatial information (SI) \cite{SI-TI}, temporal information (TI) \cite{SI-TI}, and colorfulness (CF) \cite{CF}, as well as semantic-aware deep learning models like ResNet50 and VGG16, to predict cross-over bitrates between consecutive resolutions. They found the Extra-Trees regressor to perform the best.

In \cite{Perceptually-Aware-Per-Title-Encoding-for-Adaptive-Video-Streaming}, the authors modeled VMAF of a compressed video as a linear regression of bitrate and DCT-based texture energy features. They reported a good correlation against VMAF scores with the coefficients of the linear regression being allowed to vary with resolution. The authors later used this quality prediction model to construct per-shot bitrate ladders. The same authors extended their work in \cite{Just-Noticeable-Difference-aware-Per-Scene-Bitrate-laddering-for-Adaptive-Video-Streaming} by predicting VMAF, CRFs, and JND thresholds on compressed videos using the same DCT-based energy features, luminescence, and bitrate. The number of predicted encoding settings was further reduced during bitrate ladder construction using JND thresholds predicted using GLCM features, bitstream features, etc. The authors in \cite{Ensemble-Learning-for-Efficient-VVC-Bitrate-Ladder-Prediction} combined a classifier that predicts optimal resolution with a regressor that predicts cross-over bitrates, to create an ensemble aggregator that constructs bitrate ladders. The authors also used low-level spatio-temporal features like GLCM and TC. Instead of designing models to construct optimal Pareto-fronts, the authors of \cite{Convex-Hull-Prediction-for-Adaptive-Video-Streaming-by-Recurrent-Learning} used deep learning models to predict points on the Pareto-Front, by modeling the problem as a multi-label classification problem. The authors deployed a deep learning model with Conv-GRU units to extract spatio-temporal features on the videos before compression. Techniques like incremental learning to achieve tractable memory footprints, and transfer learning to analyze wide ranges of content complexities were used to augment training of their deep learning models.

In earlier preliminary work \cite{Bitrate-Ladder-Construction-using-Visual-Information-Fidelity}, we trained an Extra-Trees regressor on multiple Visual Information Fidelity (VIF) \cite{VIF} feature sets extracted over different scales and subbands of source videos. In addition to the VIF features, we used metadata like bitrates, and the widths, and heights of the compressed videos to predict the VMAF of the corresponding compressed videos. We used these models to construct per-shot bitrate ladders. One advantage of using VIF feature sets is that they are readily available when using widely deployed full-reference (FR) quality estimation model VMAF.

\section{Method}
\label{sec:method}
Next, we describe the problem formulation, training and inference procedures and the features employed in our bitrate and quality ladder model.

\begin{figure}
	\centering
	\includegraphics[width=0.75\columnwidth]{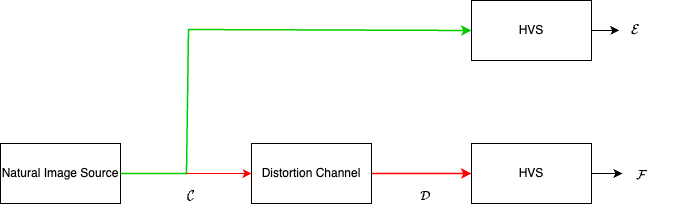}
	\caption{An illustration showing differences between VIF feature extraction in our method and VIF quality score estimation in full-reference video quality assessment. The block diagram is similar to a figure in \cite{VIF}. The green and red lines in the block diagram together represent the VIF features used for quality prediction and the green lines represent the VIF features used in our method.}
	\label{fig:FR-VIF}
\end{figure}

\begin{figure*}
	\centering
	\includegraphics[width=0.85\textwidth]{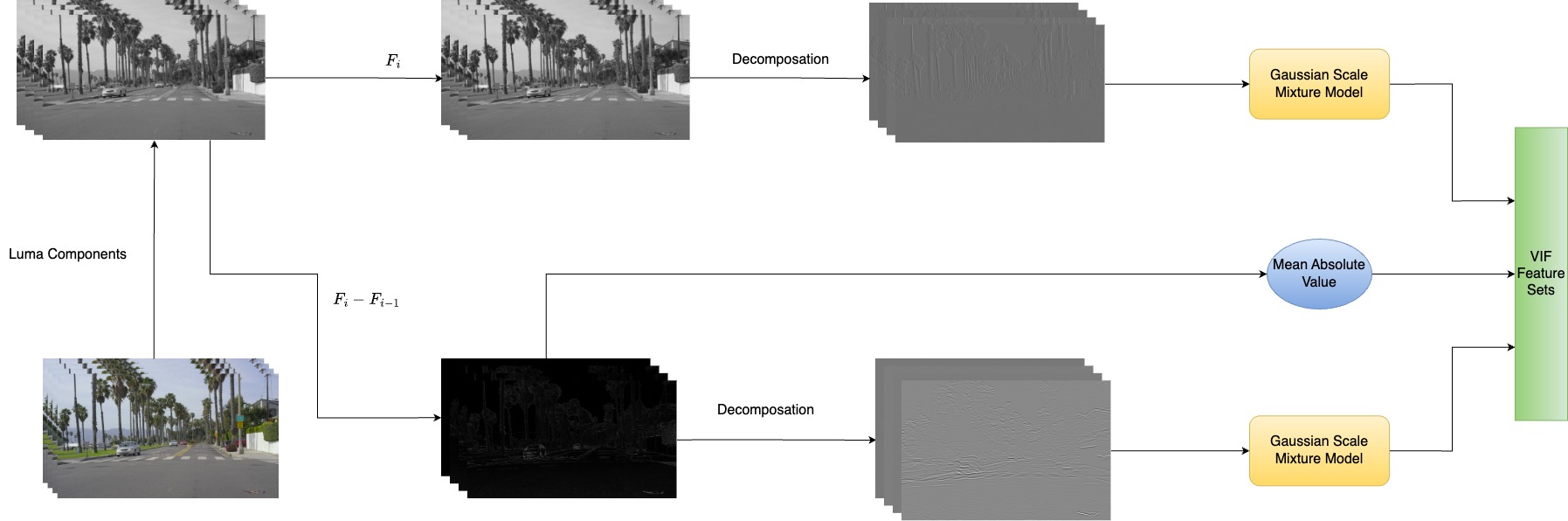}
	\caption{VIF Feature Extraction}
	\label{fig:VIF-Feature-Extraction}
\end{figure*}

\subsection{Problem Formulation}
We formulate the problem of constructing per-shot bitrate and quality ladders as a quality and bitrate prediction problem, respectively. Unlike the fixed bitrate ladder \cite{Fixed-Bitrate-Ladder}, we aim to employ machine learning methods to construct per-shot content-gnostic bitrate and quality ladders. A convex hull or reference bitrate ladder (sampled from a convex hull) is constructed by encoding the given video across multiple resolutions and rate-control settings, then using a full-reference quality estimation model to estimate the quality of the compressed video. Here, we replace the time-consuming encoding and quality estimation procedures with machine learning models that directly predict the qualities or bitrates of videos to be compressed.

\subsubsection{Training}
Fig. \ref{fig:Proposed-Framework-Training} shows the training framework of our method. We train regressors using video features extracted from a source video, along with metadata obtained by compressing and estimating the quality of the compressed video. We use VMAF \cite{VMAF} to estimate the qualities of compressed videos. We train quality prediction regressors to predict the scaled VMAF scores ($q/100$, $q$ is VMAF) of compressed videos using video features, the logarithm of bitrate ($\log_{2}(b)$), scaled widths ($\frac{w}{3840}$), and the scaled heights ($\frac{h}{3840}$) of the compressed videos. Similarly, we train bitrate prediction regressors to predict the logarithm of bitrate ($\log_{2}(b)$) of each compressed video using video features, scaled VMAF scores ($q/100$, $q$ is VMAF), scaled widths ($\frac{w}{3840}$), and the scaled heights ($\frac{h}{3840}$) of compressed videos. It is also worth noting that, unlike regressors that predict cross-over points \cite{Content-gnostic-Bitrate-Ladder-Prediction-for-Adaptive-Video-Streaming,Efficient-Bitrate-Ladder-Construction-for-Content-Optimized-Adaptive-Video-Streaming, Benchmarking-Learning-based-Bitrate-Ladder-Prediction-Methods-for-Adaptive-Video-Streaming}, the quality or bitrate prediction regressors provide supplementary information regarding the approximate bitrates or qualities of the compressed videos. We trained multiple regressors, including Extra-Trees \cite{Extra-Trees}, XG-Boost \cite{XGBoost}, and RandomForests \cite{Random-Forest}, on each task in our experiments. Machine learning models were trained separately on each feature set/approach. We found that the Extra-Trees regressor consistently delivered the best results on all our feature sets on almost every task.

\subsubsection{Inference}
Fig. \ref{fig:Proposed-Framework-Testing} shows the inference framework of our proposed method. A bitrate ladder contains steps of bitrate and the corresponding optimal resolution for encoding at that bitrate step. When constructing the bitrate ladders, we selected the following bitrates (in kbps) as steps: 500, 1000, 2000, 3000, 4000, 5000, 6000, 7000, 8000, 9000, 10500, 12000, 15000. Similarly, a quality ladder contains quality steps based on the full-reference model is employed and the corresponding optimal resolution for encoding at that quality step. When constructing quality ladders, we selected the following VMAF scores as steps: 25, 35, 45, 50, 55, 60, 65, 70, 75, 80, 85, 90, 92.5. To predict the per-shot bitrate ladder of a given video, we employ the quality prediction regressors to predict quality scores of videos compressed at each bitrate step across multiple resolutions. The resolution that yields the highest prediction is considered as optimal at the corresponding bitrate step. Conversely, when predicting the per-shot quality ladder, we employ bitrate prediction regressors to predict the bitrates of compressed videos at each quality step across multiple resolutions. The resolution that yields the lowest prediction is considered optimal at the corresponding quality step. In both these scenarios, we construct per-shot bitrate/quality ladders without employing compression or quality estimation modules.

\subsection{VIF Feature Sets}
Visual Multimethod Assessment Fusion (VMAF) \cite{VMAF} is a widely used full-reference video quality assessment (VQA) model that has demonstrated excellent correlations against human judgments of video quality. VMAF uses a Support Vector regressor (SVR) model to fuse spatial features from VIF \cite{VIF}, the Detail Loss Metric (DLM) \cite{DLM}, and a simple temporal frame-difference feature computed as the average absolute luminance differences between adjacent frames. VIF \cite{VIF} is a full-reference image quality assessment (IQA) that predicts the information that could ideally be extracted by visual neurons from the reference image relative to the loss of information from distortion. It uses a Gaussian scale mixture image model expressed in the wavelet domain.

Fig. \ref{fig:FR-VIF} shows the differences between the VIF feature extraction in our model, and VIF quality estimation as used in full-reference video quality assessment engines. The block diagram is similar to a figure in \cite{VIF}. The green and red lines in the block diagram represent the paths we considered and excluded, respectively. Unlike during quality assessment, we only compute the information content of the source video as the mutual information between the input and output of the HVS channel. Fig. \ref{fig:VIF-Feature-Extraction} shows the VIF feature extraction procedure. VIF feature sets consist of features computed at multiple scales and subbands of frames from the reference video, as well as luminance differences between adjacent frames. VIF features are defined as follows:
\begin{align}
    \mathcal{C} = \mathcal{S} . \mathcal{U} = \{S_{i}.\vec{U}_{i}: i \in I\} \\
    \mathcal{E} = \mathcal{C} + \mathcal{N}
\end{align}
\begin{align}
    I(\vec{C}^N;\vec{E}^N | s^N) = \sum_{j=1}^{N}\sum_{i=1}^{N} I(\vec{C}_{i};\vec{E}_{j} | \vec{C}^{i-1}, \vec{E}^{j-1}, s^N) \\
	I(\vec{C}^N;\vec{E}^N | S^N = s^N) = \sum_{i=1}^{N} I(\vec{C}_{i};\vec{E}_{i} | s_{i}) \\
    I(\vec{C}^N;\vec{E}^N | s^N) = \frac{1}{2}\sum_{i=1}^{N} \log_{2}(\frac{|s_{i}^2\mathbf{C_{U}} + \sigma_{n}^{2}\mathbf{I}|}{|\sigma_{n}^{2}\mathbf{I}|}), \label{eqn:vif0}
\end{align}

\noindent where $\mathcal{C} = \{\vec{C}_{i}: i \in I\}$ is a random field (RF) representing a subband of the reference image, $\mathcal{S} = \{\vec{S}_{i}: i \in I\}$ is a RF of positive scalars, and $\mathcal{U} = \{\vec{U}_{i}: i \in I\}$ is a Gaussian RF having mean zero and covariance $\mathbf{C}_{U}$. $\vec{C}_{i}$ and $\vec{U}_{i}$ are M-dimensional vectors and $\vec{U}_{i}$ is independent of $\vec{U}_{j}$ when $i \neq j$. $\mathcal{E} = \{\vec{E}_{i}: i \in I\}$ is the output of the neural model and $\mathcal{N} = \{\vec{N}_{i}: i \in I\}$ models noise and uncertainty in the wavelet domain as a multivariate Gaussian having mean zero and covariance $\mathbf{C}_{N} = \sigma_{n}^2\mathbf{I}$. $I(\vec{C}^N;\vec{E}^N | s^N)$ represents the information that could ideally be extracted from a particular subband in the image. $\vec{s}^N$ is a realization of $S^N = (S_{1}, S_{2}, \dots, S_{N})$ on a particular reference image, which can be thought of as model parameters associated with it. Matrix factorization of \eqref{eqn:vif0} yields:
\begin{align}
	I(\vec{C}^N;\vec{E}^N | s^N) = \frac{1}{2}\sum_{i=1}^{N}\sum_{j=1}^{M} \log_{2}(1 + \frac{s_{i}^2\lambda_{j}}{\sigma_{n}^{2}}).
\end{align}
We calculate all the above features over four scales, each having two subbands, using $M = 9$. We consider $I_{k,b}^{j}$ to represent mutual information along the $j^{th}$ eigenvector of the $b^{th}$ subband at the $k^{th}$ scale, where $j \in \{1,2,...,M\}$, $b \in \{1,2\}$, and $k \in \{1,2,3,4\}$:
\begin{align}
	I_{k,b}^{j} = \frac{1}{N}\sum_{i=1}^{N} \log_{2}(1 + \frac{s_{i}^2\lambda_{j}}{\sigma_{n}^{2}}) \\
	I_{k,b} = \frac{1}{N}\sum_{j=1}^{M}\sum_{i=1}^{N} \log_{2}(1 + \frac{s_{i}^2\lambda_{j}}{\sigma_{n}^{2}}) \\
	I_{k} = \frac{1}{2}\sum_{b=1}^{2}I_{k,b}.
\end{align}

The above features $I_{k}$, $I_{k,b}$, and $I_{k,b}^{j}$ are VIF features at each scale, subband, and along the direction of eigenvector, respectively. Since VIF only captures spatial characteristics, similar to VMAF, we also compute the mean absolute luminance component difference between consecutive frames. As additional temporal features, we apply the same VIF features on differences between the luminance components of consecutive frames, as was done in \cite{A-visual-information-fidelity-approach-to-video-quality-assessment}. Table \ref{table:vif-features-sets} shows the nine different feature sets used in our experiments, where $F_{i}$ represents the luminance component of the $i^{\text{th}}$ frame of a source video, and $D_{i} = F_{i} - F_{i-1}$ is the difference between luminance components of adjacent frames. We temporally pool each VIF feature by calculating their means. Hence, as depicted in Figure \ref{fig:VIF-Feature-Extraction}, the VIF features can be computed using modules already incorporated within the VMAF framework. Furthermore, since the proposed features are exclusively computed on the source video, they can be calculated without employing both compression and quality estimation.

\begin{table}
	\renewcommand{\arraystretch}{1.5}
	\centering
	\caption{List of VIF feature sets.}
	\resizebox{\columnwidth}{!}{
	\begin{tabular}{| m{8em} | m{12em} | m{6em} |} 
	\hline
	\textbf{Notation for VIF feature set} & \textbf{Features} & \textbf{No.of Features} \\ 
	\hline
	$\text{VIFF}_{1}$ & $I_{k}[F_{i}]$ & 4 \\ 
	\hline
	$\text{VIFF}_{2}$ & $I_{k,b}[F_{i}]$ & 8\\ 
	\hline
	$\text{VIFF}_{3}$ & $I_{k,b}^{j}[F_{i}]$ & 72\\ 
	\hline
	$\text{VIFF}_{4}$ & $I_{k}[F_{i}]$, $|D_{i}|$ & 5 \\ 
	\hline
	$\text{VIFF}_{5}$ & $I_{k,b}[F_{i}]$, $|D_{i}|$ & 9\\ 
	\hline
	$\text{VIFF}_{6}$ & $I_{k,b}^{j}[F_{i}]$, $|D_{i}|$ & 73\\ 
	\hline
	$\text{VIFF}_{7}$ & $I_{k}[F_{i}]$, $|D_{i}|$, $I_{k}[D_{i}]$ & 9 \\ 
	\hline
	$\text{VIFF}_{8}$ & $I_{k,b}[F_{i}]$, $|D_{i}|$, $I_{k,b}[D_{i}]$ & 17\\ 
	\hline
	$\text{VIFF}_{9}$ & $I_{k,b}^{j}[F_{i}]$, $|D_{i}|$, $I_{k,b}^{j}[D_{i}]$ & 145\\
	\hline
	\end{tabular}}
	\label{table:vif-features-sets}
\end{table}

\subsection{Low-Level Features}
\begin{table}
	\renewcommand{\arraystretch}{2}
	\normalfont
	\normalsfcodes
	\centering
	\caption{List of low-level features.}
	\resizebox{\columnwidth}{!}{
	\begin{tabular}[]{| m{4em} | m{30em} | m{4em} |}
	\hline
	\textbf{Feature} & \textbf{Formula} & \textbf{No.of Features}\\
	\hline
	GLCM & $F_{2}\{F_{1}\{correlation(GLCM)\}\}$, $F_{2}\{F_{1}\{contrast(GLCM)\}\}$, $F_{2}\{F_{1}\{energy(GLCM)\}\}$, $F_{2}\{F_{1}\{homogeneity(GLCM)\}\}$ where GLCM is calculated on blocks of size (64,64), $F_{1} = \{mean, std\}$ and $F_{2} = \{mean, std, skew, kurtosis\}$ & 32\\
	\hline
	TC & $F_{2}\{F_{1}\{Coherence\}\}$ where $F_{1} = \{mean, std, skew, kurtosis\}$ and $F_{2} = \{mean, std\}$ & 8\\
	\hline
	SI & $F_{2}\{F_{1}\{Sobel(Y)\}\}$ where $F_{1} = \{mean, std\}$ and $F_{2} = \{mean, std, skew, kurtosis\}$ & 8\\
	\hline
	TI & $F_{2}\{F_{1}\{(Y_{i+1} - Y_{i})\}\}$ where $F_{1} = \{mean, std\}$ and $F_{2} = \{mean, std, skew, kurtosis\}$ & 8\\
	\hline
	CTI & $F_{2}\{F_{1}\{Y\}\}$ where $F_{1} = \{mean, std\}$ and $F_{2} = \{mean, std, skew, kurtosis\}$ & 8\\
	\hline
	CF & $F_{2}\{(YUV)\}$ where $F_{2} = \{mean, std, skew, kurtosis\}$ & 4\\
	\hline
	CI & $F_{2}\{F_{1}\{U\}\}$, $F_{2}\{W_{R} \times F_{1}\{V\}\}$ where $W_{R} = 5$, $F_{1} = \{mean, std\}$ and $F_{2} = \{mean, std, skew, kurtosis\}$ & 16\\
	\hline
	DCT-Texture & $F_{2}\{E_{Y}\}$, $F_{2}\{h_{Y}\}$, $F_{2}\{L_{Y}\}$, $F_{2}\{E_{U}\}$, $F_{2}\{h_{U}\}$, $F_{2}\{L_{U}\}$, $F_{2}\{E_{V}\}$, $F_{2}\{h_{V}\}$, $F_{2}\{L_{V}\}$ where $F_{2} = \{mean\}$ & 9\\
	\hline
	Bitrate-DCT-Texture & $\log_{2}\left[\sqrt{\frac{F_{2}\{h_{Y}\}}{F_{2}\{E_{Y}\}}}\right] + 2\log_{2}(b)$, $\log_{2}\left[\sqrt{\frac{F_{2}\{h_{U}\}}{F_{2}\{E_{U}\}}}\right] + 2\log_{2}(b)$, $\log_{2}\left[\sqrt{\frac{F_{2}\{h_{V}\}}{F_{2}\{E_{V}\}}}\right] + 2\log_{2}(b)$ where $F_{2} = \{mean\}$ & 3\\
	\hline
	VMAF-DCT-Texture & $\frac{1}{2}\left(q - \log_{2}\left[\sqrt{\frac{F_{2}\{h_{Y}\}}{F_{2}\{E_{Y}\}}}\right]\right)$, $\frac{1}{2}\left(q - \log_{2}\left[\sqrt{\frac{F_{2}\{h_{U}\}}{F_{2}\{E_{U}\}}}\right]\right)$, $\frac{1}{2}\left(q - \log_{2}\left[\sqrt{\frac{F_{2}\{h_{V}\}}{F_{2}\{E_{V}\}}}\right]\right)$, where $F_{2} = \{mean\}$ & 3\\
	\hline
	\end{tabular}}
	\label{table:low-level-features}
\end{table}

As discussed in Section \ref{sec:related_works}, prior models often employ low-level features like gray-level co-occurrence matrices (GLCM) \cite{GLCM}, measures of temporal coherence (TC) \cite{TC}, normalized cross-correlations (NCC) and other content-sensitive features when modeling RQ-curves and predicting optimal ladder parameters. Traditional spatial information (SI) \cite{SI-TI} and temporal information (TI) \cite{SI-TI} defined by simple spatial or temporal differencing, respectively, have commonly been used to measure video complexity. However, other authors \cite{Towards-Perceptually-Optimized-Compression-Of-User-Generated-Content}, \cite{VCA}, have shown that chrominance features and DCT-based texture energy features proved more effective when modeling RQ characteristics. 

Our study incorporates features derived from an array of prior work on video coding, to develop a comprehensive set of low-level features. We utilize GLCM and TC features, as in \cite{Predicting-Video-Rate-Distortion-Curves-using-Textural-Features, Study-of-compression-statistics-and-prediction-of-rate-distortion-curves-for-video-texture, Content-gnostic-Bitrate-Ladder-Prediction-for-Adaptive-Video-Streaming, Efficient-Bitrate-Ladder-Construction-for-Content-Optimized-Adaptive-Video-Streaming, VMAF-based-Bitrate-Ladder-Estimation-for-Adaptive-Streaming, Benchmarking-Learning-based-Bitrate-Ladder-Prediction-Methods-for-Adaptive-Video-Streaming}, and SI, TI, contrast information (CTI) \cite{CTI-CI}, chrominance information (CI) \cite{CTI-CI}, and colorfulness (CF) \cite{CF} as employed in \cite{Towards-Perceptually-Optimized-Compression-Of-User-Generated-Content}. We also compute DCT-based texture energy features as introduced in \cite{Perceptually-Aware-Per-Title-Encoding-for-Adaptive-Video-Streaming, VCA}, and extended to chroma components in \cite{Just-Noticeable-Difference-aware-Per-Scene-Bitrate-laddering-for-Adaptive-Video-Streaming}.

The authors of \cite{Perceptually-Aware-Per-Title-Encoding-for-Adaptive-Video-Streaming}, devised a bitrate dependent feature that combines spatial and temporal energy features of luminance: $\log_{2}\left[\sqrt{\frac{h_{Y}}{E_{Y}}}\right] + 2\log_{2}(b)$, ($E_{Y}$ is spatial energy, $h_{Y}$ is temporal energy, and $b$ is bitrate) which they used to predict VMAF scores. In addition to the aforementioned feature, we compute features extending computation on the same features expressed on chroma components. We also devised quality dependent features similar to bitrate dependent features, but using VMAF scores instead of bitrates. Table \ref{table:low-level-features} shows the list of features we deploy in our work where $GLCM$ is gray-level co-occurrence matrix, $Coherence$ is measure of temporal coherence, $Y$ is luminance, $U$ and $V$ are chroma components, $E_{Y}$, $E_{U}$, $E_{V}$ are spatial energy components, $h_{Y}$, $h_{U}$, $h_{V}$ are temporal energy components, $L_{Y}$, $L_{U}$, $L_{V}$ are luminescence components, $b$ is bitrate, and $q$ is quality. $F_{1}$ operation denotes spatial feature pooling and $F_{2}$ denotes temporal feature pooling.

\begin{figure}
	\centering
	\begin{subfigure}[b]{0.49\columnwidth}
		\centering
		\includegraphics[width=\columnwidth]{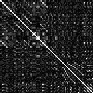}
		\caption{$|\text{PLCC}|$}
		\label{correlation-1}
	\end{subfigure}
	\hfill
	\begin{subfigure}[b]{0.49\columnwidth}
		\centering
		\includegraphics[width=\columnwidth]{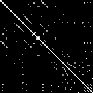}
		\caption{$|\text{PLCC}| \geq 0.5$}
		\label{correlation-2}
	\end{subfigure}
	\caption{Correlation among low-level features.}
	\label{fig:correlation}
\end{figure}

\begin{figure}
	\centering
	\begin{subfigure}[b]{0.49\columnwidth}
		\centering
		\includegraphics[width=\columnwidth]{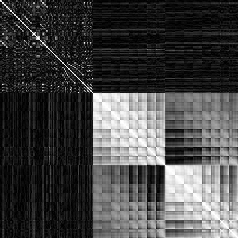}
		\caption{$|\text{PLCC}|$}
		\label{correlation-3}
	\end{subfigure}
	\hfill
	\begin{subfigure}[b]{0.49\columnwidth}
		\centering
		\includegraphics[width=\columnwidth]{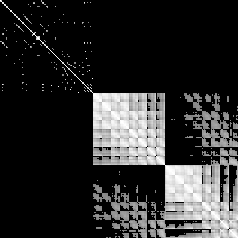}
		\caption{$|\text{PLCC}| \geq 0.5$}
		\label{correlation-4}
	\end{subfigure}
	\caption{Correlation among low-level features and VIF features ($\text{VIFF}_{9}$).}
	\label{fig:cross-correlation}
\end{figure}

Similar to VIF features, low-level features are computed only on the source video, thereby eliminating the need for compression and quality estimation. To demonstrate the efficacy of our comprehensive set of low-level features, we plot the correlations among these features. Figure \ref{correlation-1} presents a $93 \times 93$ image matrix representing the absolute PLCC values among all low-level features, excluding bitrate and quality dependent DCT features. Figure \ref{correlation-2} shows the same PLCC values, but only displaying correlations having absolute values $\geq 0.5$, while zeroing out the rest. It may observed that the low-level features we use show very little correlation with each other, with cross-correlation values between most of the features below $0.5$. This indicates that the considered features are complementary to each other, and can be used to build a robust model. Similarly, Fig. \ref{correlation-3} and Fig. \ref{correlation-4} show the correlations among low-level features and VIF features ($\text{VIFF}_{9}$). It may be observed that the considered set of low-level features is little correlated with VIF features ($ < 0.5$). Hence, we may conclude that the considered set of content-aware low-level features and quality-aware VIF features supply only weakly uncorrelated information. The correlation among VIF features ($\text{VIFF}_{9}$) is also evident from Figures \ref{correlation-3} and \ref{correlation-4}. Although not as effective as the cross-correlation among low-level features, VIF features exhibit less cross-correlation, particularly between spatial and temporal components.

\section{Experiments and Results}
\label{sec:experiments-and-results}
\subsection{Dataset}
\begin{figure}
    \centering
    \includegraphics[width=0.9\linewidth]{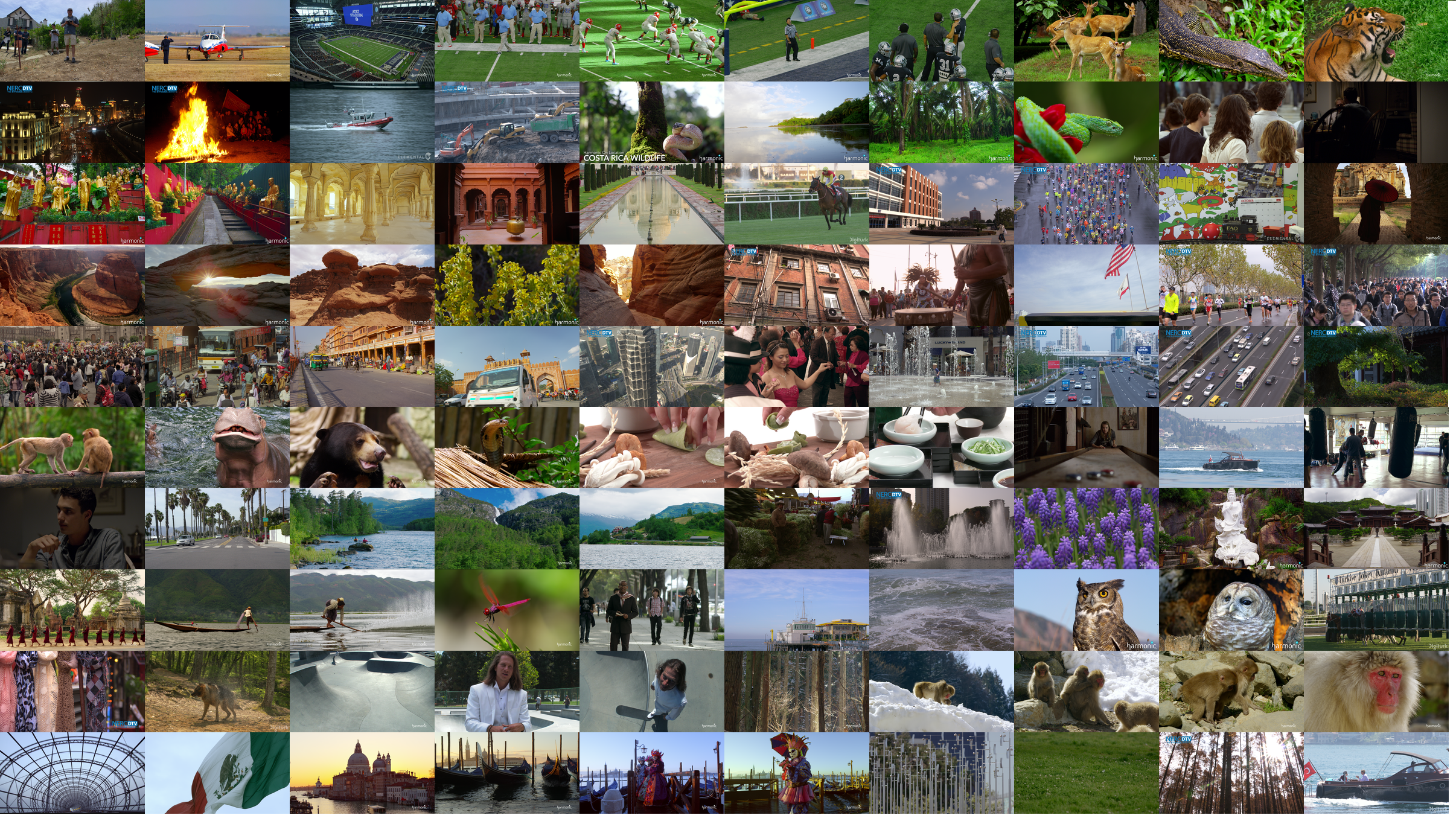}
    \caption{Sample frames from each video taken from the BVT-100 4K dataset \cite{Efficient-Bitrate-Ladder-Construction-for-Content-Optimized-Adaptive-Video-Streaming}.}
    \label{fig:BVT-100_4K}
\end{figure}

Our video data was drawn from the BVT-100 4K dataset used in \cite{Content-gnostic-Bitrate-Ladder-Prediction-for-Adaptive-Video-Streaming},\cite{Efficient-Bitrate-Ladder-Construction-for-Content-Optimized-Adaptive-Video-Streaming}. This dataset consists of 100 video sequences derived from various public sources including Netflix Chimera \cite{Netflix-Chimera}, Ultra Video Group \cite{UVG}, Harmonic Inc. \cite{HarmonicFootage}, SJTU \cite{SJTU}, and AWS Elemental \cite{AWSElemental}. All of the video sequences were converted to 4:2:0 chroma subsampling, spatially cropped to UHD (3840$\times$2160 pixels), and temporally clipped to 64 frames. Each video was constrained to contain a single scene (without scene cuts). The majority of the test videos have frame rates of 60 fps and bit depths of 10 bits/pixel of luma and chroma. Fig. \ref{fig:BVT-100_4K} shows sample frames of the videos comprising in the dataset.

\subsection{Experiment Settings}
\begin{figure*}
    \centering
    \begin{subfigure}[b]{0.325\linewidth}
    \centering
    \includegraphics[width=\linewidth]{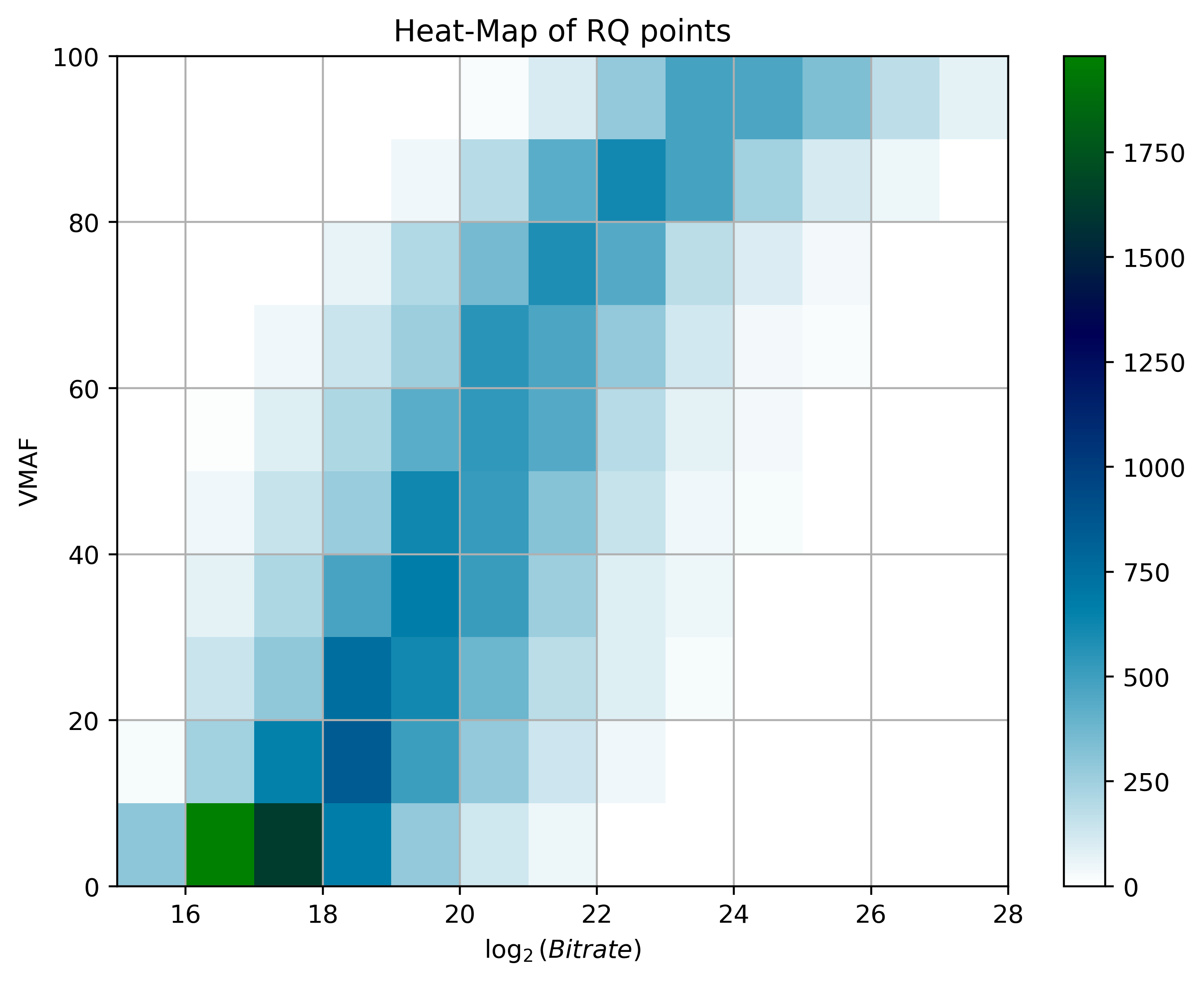}
    \caption{Heat-map of RQ points over 8 resolutions and 31 CRFs.}
    \label{fig:RQ_Histogram_init}
    \end{subfigure}
    \begin{subfigure}[b]{0.325\linewidth}
    \centering
    \includegraphics[width=\linewidth]{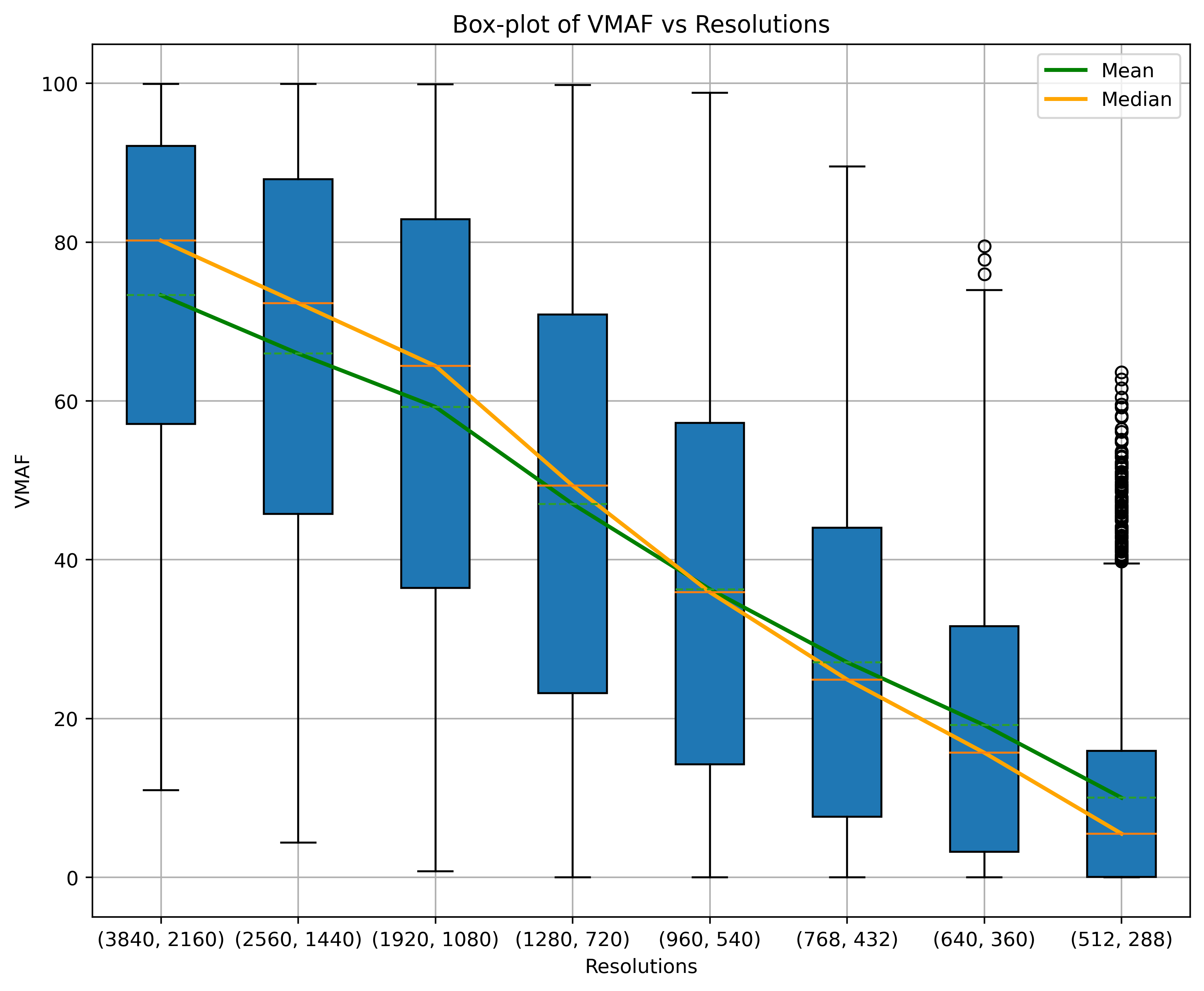}
    \caption{Box-plot of VMAF vs spatial resolution.\\\hfill}
    \label{fig:Box_Plot_Resolutions_VMAF_init}
    \end{subfigure}
    \begin{subfigure}[b]{0.325\linewidth}
    \centering
    \includegraphics[width=\linewidth]{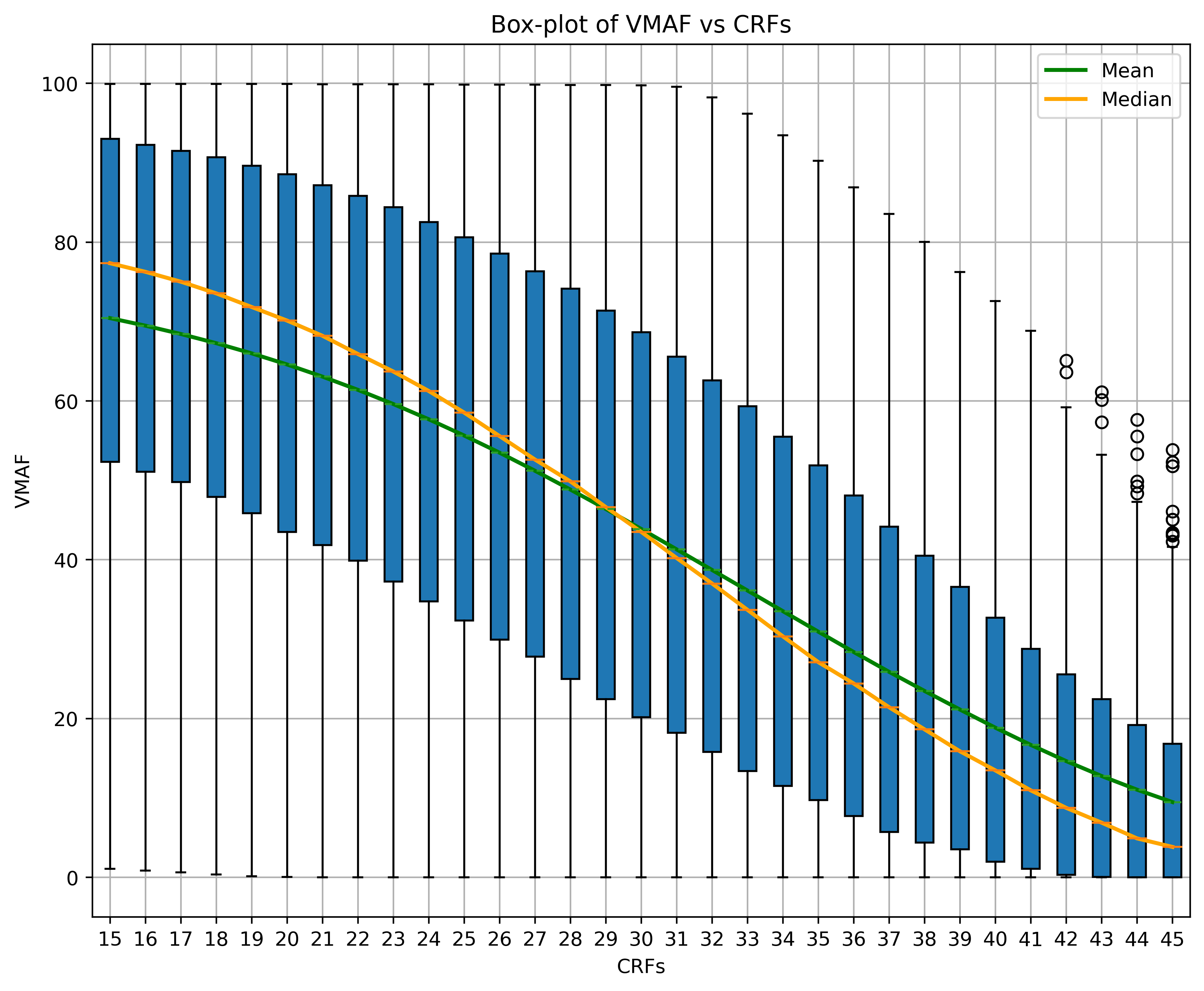}
    \caption{Box-plot of VMAF vs CRFs.\\\hfill}
    \label{fig:Box_Plot_CRFs_VMAF_init}
    \end{subfigure}
    \caption{Distributions of rate-quality points over 8 spatial resolutions and 31 CRFs of all videos in the BVT-100 4K dataset.}
    \label{fig:Distribution}
\end{figure*}

We used \textit{ffmpeg} to perform compression and quality estimation. The videos were compressed using the \textbf{libx265} codec with the \textbf{medium} preset. We used VMAF \cite{VMAF} to predict the perceptual qualities of the compressed videos. VMAF has been shown to exhibit higher correlations with human judgments than PSNR. We computed VMAF after upscaling the compressed video to its original resolution (3840$\times$2160). We used the Lanczos interpolation to conduct spatial upscaling and downscaling (from 2160p). We considered eight different resolutions ranging from 2160p to 288p: 3840$\times$2160, 2560$\times$1440, 1920$\times$1080, 1280$\times$720, 960$\times$540, 768$\times$432, 640$\times$360, and 512$\times$288 all of which have aspect ratio of about 16 $:$ 9. We implemented constant-quality encoding by adjusting CRFs, aiming to maintain consistent visual quality while maximizing compression efficiency. We sampled \textbf{libx265} CRF values ranging from 15 to 45 (inclusive).

To understand the distributions of rate-quality points for various bitrate and VMAF ranges, we plotted a heat map of rate-quality points for all videos compressed using the initial set of 8 resolutions and 31 CRFs, as shown in Fig. \ref{fig:RQ_Histogram_init}. To further investigate the role of each considered resolution and CRF, we plotted box plots of the distribution of VMAF for each resolution and CRF pair, as shown in Fig. \ref{fig:Box_Plot_Resolutions_VMAF_init} and Fig. \ref{fig:Box_Plot_CRFs_VMAF_init}, respectively. It may be observed that the distribution of VMAF scores was negatively skewed for higher resolutions and smaller CRFs, while VMAF scores are positively skewed for lower resolutions and larger CRFs. The majority of the rate-quality points at resolutions 640$\times$360 and 512$\times$288 and CRFs greater than 41 yielded VMAF scores less than 15. From the curves of median and mean in Fig. \ref{fig:Box_Plot_CRFs_VMAF_init}, it may also be observed that for large and small CRF values, the relative changes of the distribution of VMAF scores were slowly decreasing, with the relative change being the highest at intermediate CRFs.

Considering these factors, we focused our study on rate-quality points falling within a VMAF score range of 15 to 95 (inclusive). This range excludes VMAF regions where the quality change was either imperceptible, or the perceptual quality of the video was insufficient to allow for a satisfactory visual experience. This adjustment helps to mitigate dataset bias, given that a significant proportion of RQ points have low VMAF scores, as depicted in Fig. \ref{fig:RQ_Histogram_init}. It also leads to imbalances in the training dataset, due to the varying number of points that meet the considered constraints across videos. We used the following experimental settings throughout:
\begin{itemize}
    \item \textbf{Codec, Preset}: libx265, medium
    \item \textbf{Resolutions}: 3840$\times$2160, 2560$\times$1440, 1920$\times$1080, 1280$\times$720, 960$\times$540, and 768$\times$432
    \item \textbf{CRFs}: 16 to 35 (inclusive), 35 to 41 (inclusive) with a skip of 2
    \item \textbf{Constraints}: 15 $\leq$ VMAF $\leq$ 95
\end{itemize}

\begin{figure}
    \centering
    \begin{subfigure}[b]{0.49\linewidth}
    \centering
    \includegraphics[width=\linewidth]{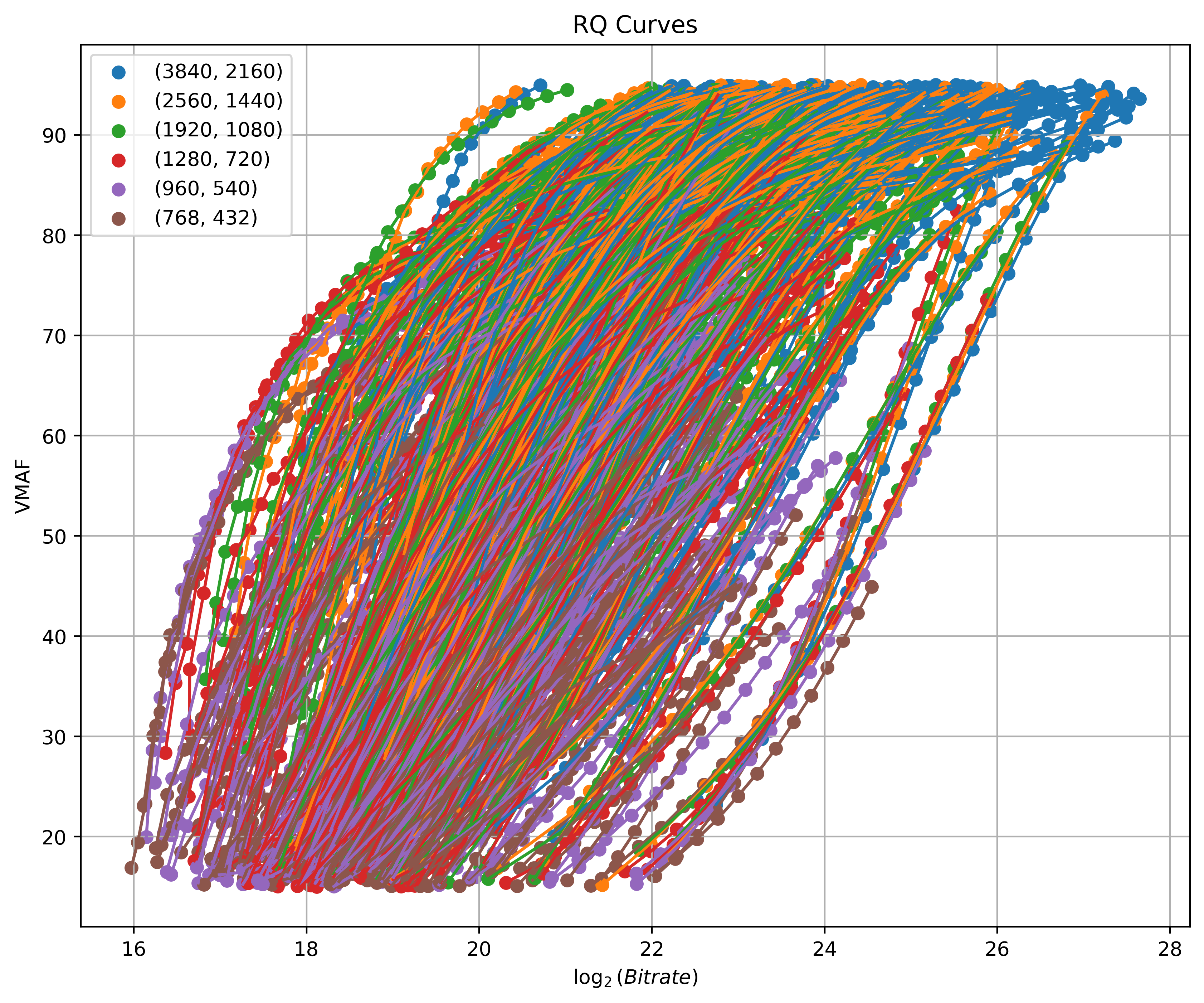}
    \caption{}
    \label{fig:rq_curves}
    \end{subfigure}
    \begin{subfigure}[b]{0.49\linewidth}
    \centering
    \includegraphics[width=\linewidth]{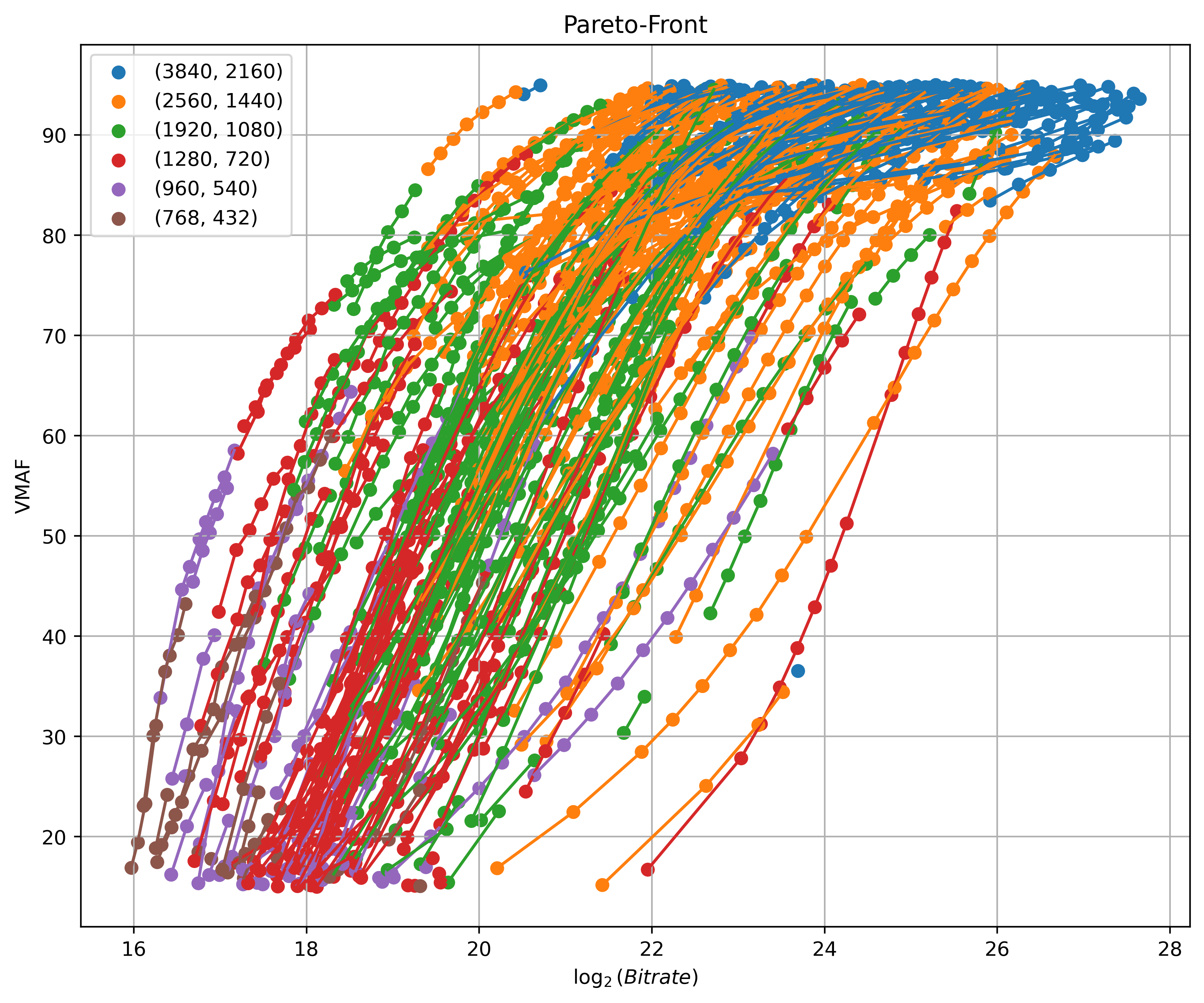}
    \caption{}
    \label{fig:pf_curves}
    \end{subfigure}
    \caption{Plots of (a) Rate-Quality curves and (b) Pareto-Fronts using exhaustive encoding, all videos in the dataset.}
    \label{fig:Curves}
\end{figure}

Fig. \ref{fig:Curves} shows the RQ curves and Pareto-Fronts constructed using exhaustive encoding of all the videos in the dataset using the experimental settings given above. One may observe great diversity among the RQ curves of videos having different resolutions, with different convexities, ranges of bitrates, and quality scores. The Pareto-fronts also display notable variety across resolutions and bitrates.

We constructed the experimental dataset by compressing all of the videos using all the encoding settings mentioned previously. We divided the dataset into non-overlapping training, validation, and test datasets containing 70, 10, and 20 videos respectively. We ensured that there were no video titles shared between these sets to avoid content learning. The videos in the validation and test datasets were considered together during evaluation, thereby yielding a larger sample size, and these videos were not part of the training data. 

\begin{table}
	\centering
	\caption{Low-level feature sets.}
	\renewcommand{\arraystretch}{1.75}
	\resizebox{\columnwidth}{!}{
	\begin{tabular}{| m{8em} | m{18em} | m{6em} |} 
	\hline
	\textbf{Notation for low-level feature set} & \textbf{Features} & \textbf{No.of Features} \\ 
	\hline
	$\text{LLF}_{1}$ & GLCM, TC, SI, TI, CTI, CF, CI, DCT-Texture & 93 \\ 
	\hline
	$\text{LLF}_{2}$ & GLCM, TC, SI, TI, CTI, CF, CI, DCT-Texture, Bitrate-DCT-Texture & 96\\ 
	\hline
	$\text{LLF}_{3}$ & GLCM, TC, SI, TI, CTI, CF, CI, DCT-Texture, VMAF-DCT-Texture & 96\\ 
	\hline
	\end{tabular}}
	\label{table:low-level-features-sets}
\end{table}

\subsection{Predicting Cross-Over Bitrates or VMAF using Low-Level Features}
Cross-over points between two resolutions are characterized as intersection points on the RQ curves, marking transitions from a lower to a higher resolution that offers superior quality. This intersection is defined either by a pair of QPs \cite{Content-gnostic-Bitrate-Ladder-Prediction-for-Adaptive-Video-Streaming, Efficient-Bitrate-Ladder-Construction-for-Content-Optimized-Adaptive-Video-Streaming}, one for each resolution, or by the corresponding bitrate to that intersection \cite{Benchmarking-Learning-based-Bitrate-Ladder-Prediction-Methods-for-Adaptive-Video-Streaming, Ensemble-Learning-for-Efficient-VVC-Bitrate-Ladder-Prediction}. These cross-over points are predicted by extracting features from uncompressed videos, and are then used to construct per-shot bitrate ladders. 

\begin{figure}
	\centering
	\caption{Plots of (a) Rate-Quality curves and Pareto-Fronts constructed for `pierseaside-scene1', using (b) exhaustive encoding, (c) cross-over bitrates, and (d) VMAF cross-over points.}
	\begin{subfigure}[b]{0.485\linewidth}
	\centering
	\includegraphics[width=\linewidth]{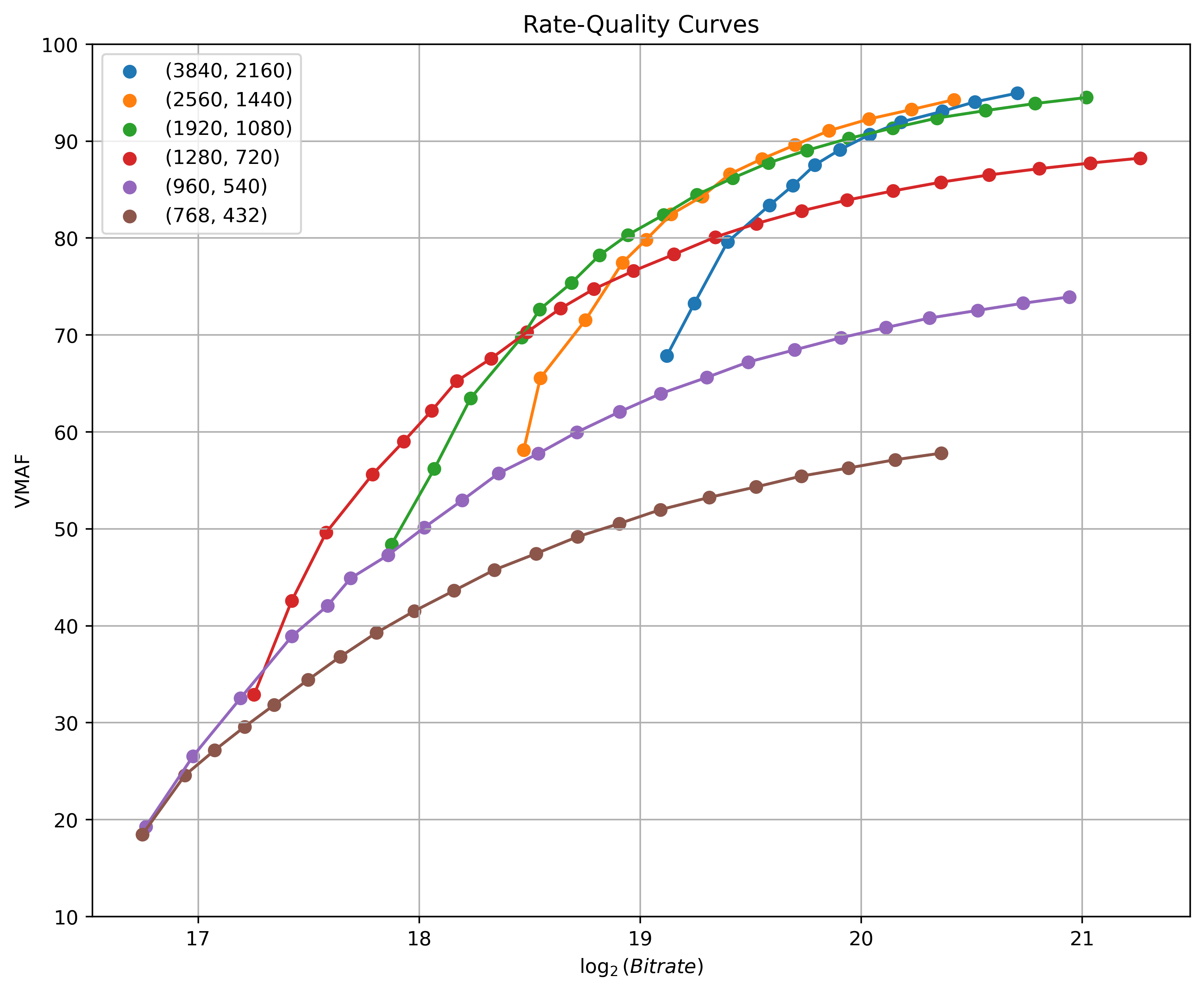}
	\caption{}
	\label{fig:rq}
	\end{subfigure}
	\begin{subfigure}[b]{0.485\linewidth}
	\centering
	\includegraphics[width=\linewidth]{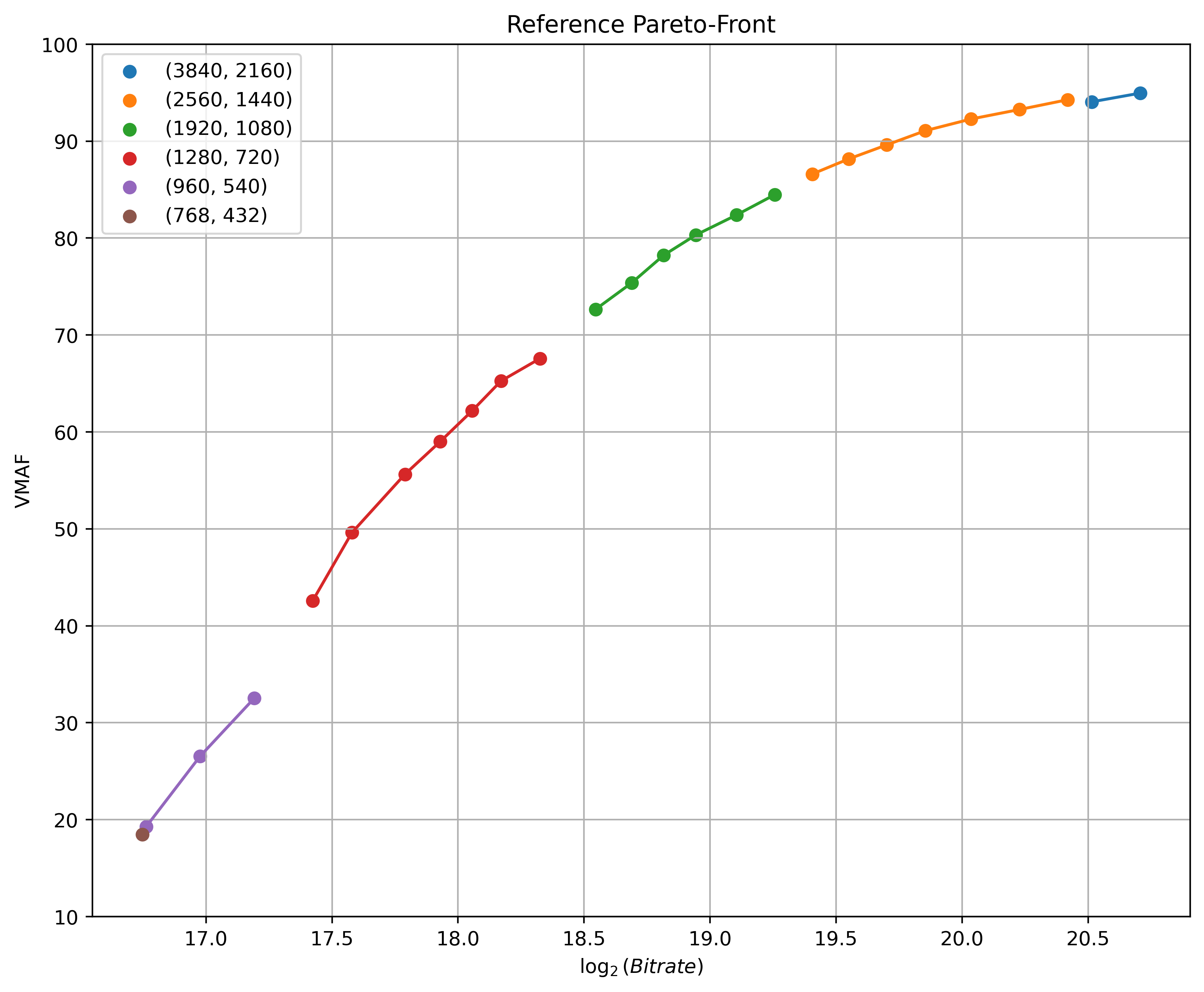}
	\caption{}
	\label{fig:pf}
	\end{subfigure}
	\begin{subfigure}[b]{0.485\linewidth}
	\centering
	\includegraphics[width=\linewidth]{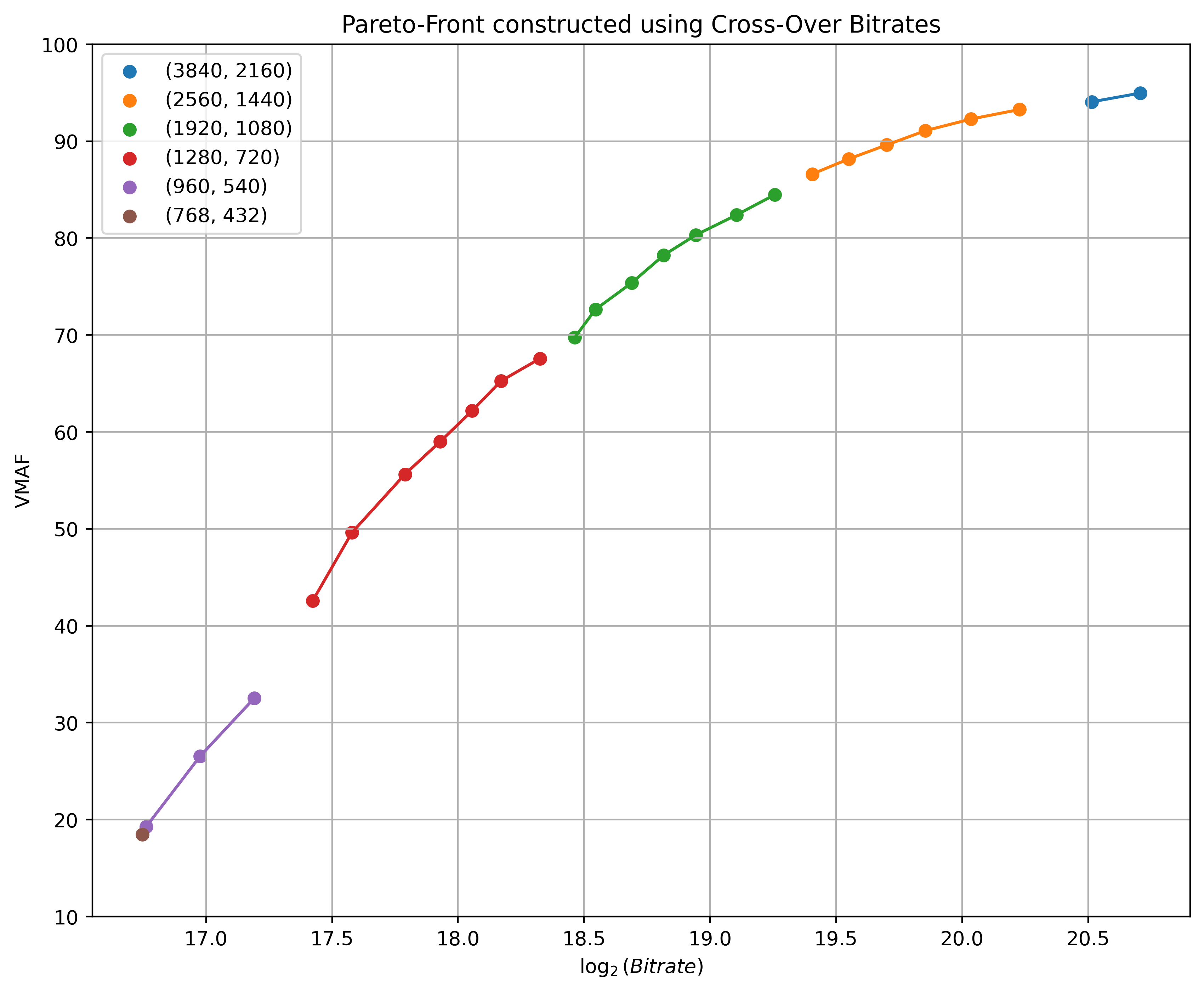}
	\caption{}
	\label{fig:cob}
	\end{subfigure}
	\begin{subfigure}[b]{0.485\linewidth}
	\centering
	\includegraphics[width=\linewidth]{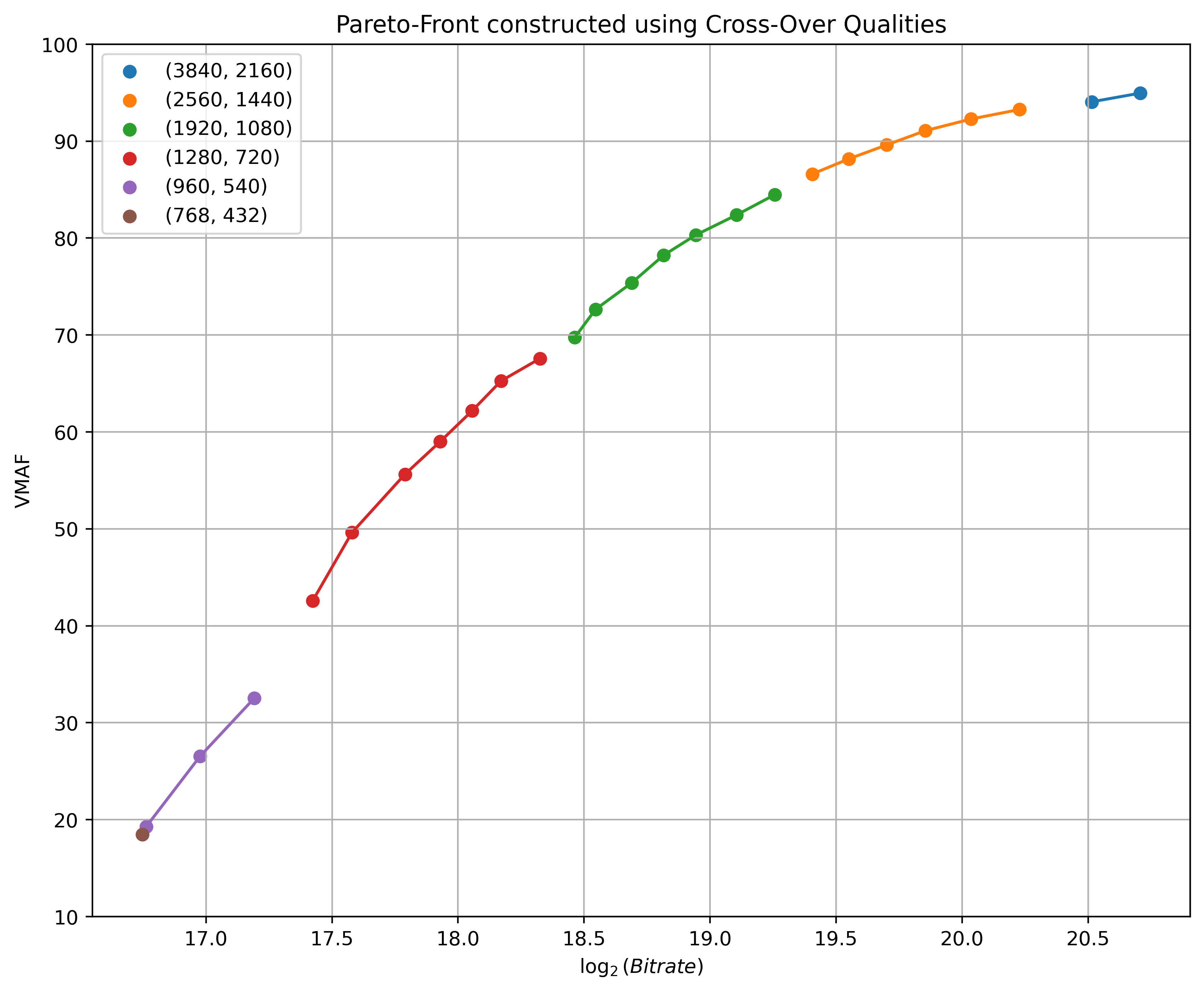}
	\caption{}
	\label{fig:coq}
	\end{subfigure}
	\label{fig:Example}
\end{figure}

Similar to bitrate cross-over points, we designed a method using VMAF cross-over points. The VMAF cross-over points are the intersections of the RQ curves at each resolution where the predicted perceptual qualities are the same. Similar to cross-over bitrates, we predict VMAF cross-over points by extracting features from the source videos, then using them to construct per-shot quality ladders. Fig. \ref{fig:Example} shows an example of rate-quality curves and Pareto-Fronts constructed using exhaustive encoding, cross-over bitrates, and VMAF cross-over points on a video sample in the dataset.

We compared our proposed methods against a technique that constructs a bitrate ladder using cross-over bitrates. After selecting six different spatial resolutions, we calculated five cross-over bitrates between adjacent resolutions. We used the low-level feature set $\text{LLF}_{1}$ defined in Table \ref{table:low-level-features-sets} in our experiments. The considered set of features is greater in number than the features employed in \cite{Benchmarking-Learning-based-Bitrate-Ladder-Prediction-Methods-for-Adaptive-Video-Streaming}. We also used predicted cross-over bitrates between higher resolutions as additional features when predicting cross-over bitrates between lower resolutions. Since our training dataset contains only 70 videos, to avoid the curse of dimensionality, we used recursive feature elimination (RFE) to select nine features that are used to predict each cross-over bitrate. Later, we use the predicted cross-over bitrates to construct per-shot bitrate ladders. As mentioned earlier, we also learned the perceptual counterpart models, \textit{viz.}, that predict VMAF cross-over points using low-level features. Similarly, we applied RFE on the set of low-level features $\text{LLF}_{1}$, to select nine features predictive of VMAF cross-over points, and we later employ these to construct per-shot quality ladders.

Table \ref{table:CO-Results} measures and compares the performances of the cross-over point prediction models defined in this way. We observed better correlations between predicted and true cross-over bitrates/VMAF at intermediate resolutions than at extreme resolutions (2160p and 432p). This is to be expected, since our quality constraints truncate different number of points as the resolutions are varied. This leads to inconsistency when calculating cross-over points at the extreme resolutions.

\begin{table}
	\centering
	\caption{Pearson correlation coefficients between true cross-over points and cross-over points predicted using low-level features between every two consecutive resolutions on the validation and test datasets.}
	\renewcommand{\arraystretch}{1.75}
	\resizebox{\columnwidth}{!}{
	\begin{tabular}{| m{6em} | c | c | c | c | c |} 
	\hline
	\textbf{Target} & \textbf{(2160p,1440p)} & \textbf{(1440p,1080p)} & \textbf{(1080p,720p)} & \textbf{(720p,540p)} & \textbf{(540p,432p)}\\ 
	\hline
	\textbf{Cross-Over Bitrates} & 0.305 & 0.762 & 0.704 & 0.625 & 0.697\\ 
	\hline
	\textbf{Cross-Over VMAFs} & 0.593 & 0.729 & 0.752 & 0.666 & 0.206\\ 
	\hline
	\end{tabular}}
	\label{table:CO-Results}
\end{table}

\subsection{Predicting Quality or Bitrate using Low-Level features}
We compared our models against \cite{Perceptually-Aware-Per-Title-Encoding-for-Adaptive-Video-Streaming}, where the authors utilized DCT-based texture energy features to predict the quality of compressed videos. Instead of only using DCT texture energy features, we develop a more comprehensive set of features, including metadata like $\log_{2}(b)$, $\frac{w}{3840}$ and $\frac{h}{3840}$, where $b$, $w$ and $h$ are the bitrates, widths, and heights of the compressed videos, respectively. The expanded list of low-level features ($\text{LLF}_{2}$) is shown in Table \ref{table:low-level-features-sets}. The number of features considered is greater than in \cite{Perceptually-Aware-Per-Title-Encoding-for-Adaptive-Video-Streaming}. The quality prediction models learned on these features are used to construct per-shot bitrate ladders. To construct per-shot quality ladders, we developed content-dependent bitrate prediction models. These models employ the more comprehensive set of low-level features $\text{LLF}_{3}$ shown in Table \ref{table:low-level-features-sets}, along with metadata including $\frac{q}{100}$, $\frac{w}{3840}$, and $\frac{h}{3840}$ where $q$ is the VMAF score of each analyzed compressed video, to predict the bitrates of compressed videos. These experiments help us understand the feasibility of predicting the bitrates or qualities of compressed videos using only low-level features extracted from the original source videos.

Table \ref{table:LLF-Results} shows the performances of quality and bitrate prediction models using only low-level video features. It may be observed that the performances of the quality prediction models was slightly higher than that of bitrate prediction models. We found that calculating aggregate PLCC values across all resolutions yields misleading interpretations of model performance at individual resolutions. Hence, we instead report the PLCC values for each resolution. Tables \ref{table:LLF-Ablation-Results-Quality-Prediction} and \ref{table:LLF-Ablation-Results-Bitrate-Prediction} report the performances of subsets of the low-level features. It may be observed from these results that the comprehensive set of LLF features $\text{LLF}_{2}$ and $\text{LLF}_{3}$ yielded better performances than did individual subsets, with the exception of the DCT texture energy features. The performances of the models trained on DCT texture energy features was slightly better or similar to the performances of models trained on $\text{LLF}_{2}$ or $\text{LLF}_{3}$ across multiple resolutions.

\begin{table}
	\centering
	\caption{Pearson correlation coefficients between true VMAF/bitrate and VMAF/bitrate predicted using low-level features at each resolution of the validation and test datasets.}
	\renewcommand{\arraystretch}{1.75}
	\resizebox{\columnwidth}{!}{
	\begin{tabular}{| c | c | c | c | c | c | c |} 
	\hline
	\textbf{Target} & \textbf{2160p} & \textbf{1440p} & \textbf{1080p} & \textbf{720p} & \textbf{540p} & \textbf{432p}\\ 
	\hline
	\textbf{Quality} & 0.580 & 0.642 & 0.668 & 0.677 & 0.643 & 0.602\\ 
	\hline
	\textbf{Bitrate} & 0.565 & 0.598 & 0.617 & 0.626 & 0.613 & 0.595\\ 
	\hline
	\end{tabular}}
	\label{table:LLF-Results}
\end{table}

\subsection{Predicting Quality or Bitrate using Metadata}
Fixed bitrate ladders \cite{Fixed-Bitrate-Ladder} are generally designed considering a variety of content-independent video characteristics, network conditions, resolutions, and bitrates. We designed a quality prediction model that aims to predict the quality of compressed videos without using any content-dependent features. This quality prediction model is the best statistical fit to the training data based on $\log_{2}(b)$, $\frac{w}{3840}$, and $\frac{h}{3840}$. Although it appears similar to a fixed bitrate ladder, this model is dependent on the characteristics of the videos in the training dataset and the encoding settings used to compress them. We also designed counterpart bitrate prediction models that only train on $\frac{q}{100}$, $\frac{w}{3840}$, and $\frac{h}{3840}$.

We compared the performances of our models against the two metadata-based models, towards understanding the advantages conferred by using video content features. Table \ref{table:Metadata-Results} shows the performances of the bitrate and quality prediction models. The performances of bitrate and quality prediction models are similar, but worse than that of the regressors trained only on low-level features (Table \ref{table:LLF-Results}).

\begin{table}
	\centering
	\caption{Pearson correlation coefficients between true VMAF/bitrate and VMAF/bitrate predicted using metadata at each resolution of the validation and test datasets.}
	\renewcommand{\arraystretch}{1.75}
	\resizebox{\columnwidth}{!}{
	\begin{tabular}{| c | c | c | c | c | c | c |} 
	\hline
	\textbf{Target} & \textbf{2160p} & \textbf{1440p} & \textbf{1080p} & \textbf{720p} & \textbf{540p} & \textbf{432p}\\ 
	\hline
	\textbf{Quality} & 0.458 & 0.521 & 0.552 & 0.548 & 0.500 & 0.421\\ 
	\hline
	\textbf{Bitrate} & 0.475 & 0.520 & 0.535 & 0.523 & 0.480 & 0.408\\ 
	\hline
	\end{tabular}}
	\label{table:Metadata-Results}
\end{table}

\subsection{Predicting Quality or Bitrate using VIF Features}
\begin{figure}
	\centering
	\begin{subfigure}[b]{0.485\linewidth}
		\centering
		\includegraphics[width=\linewidth]{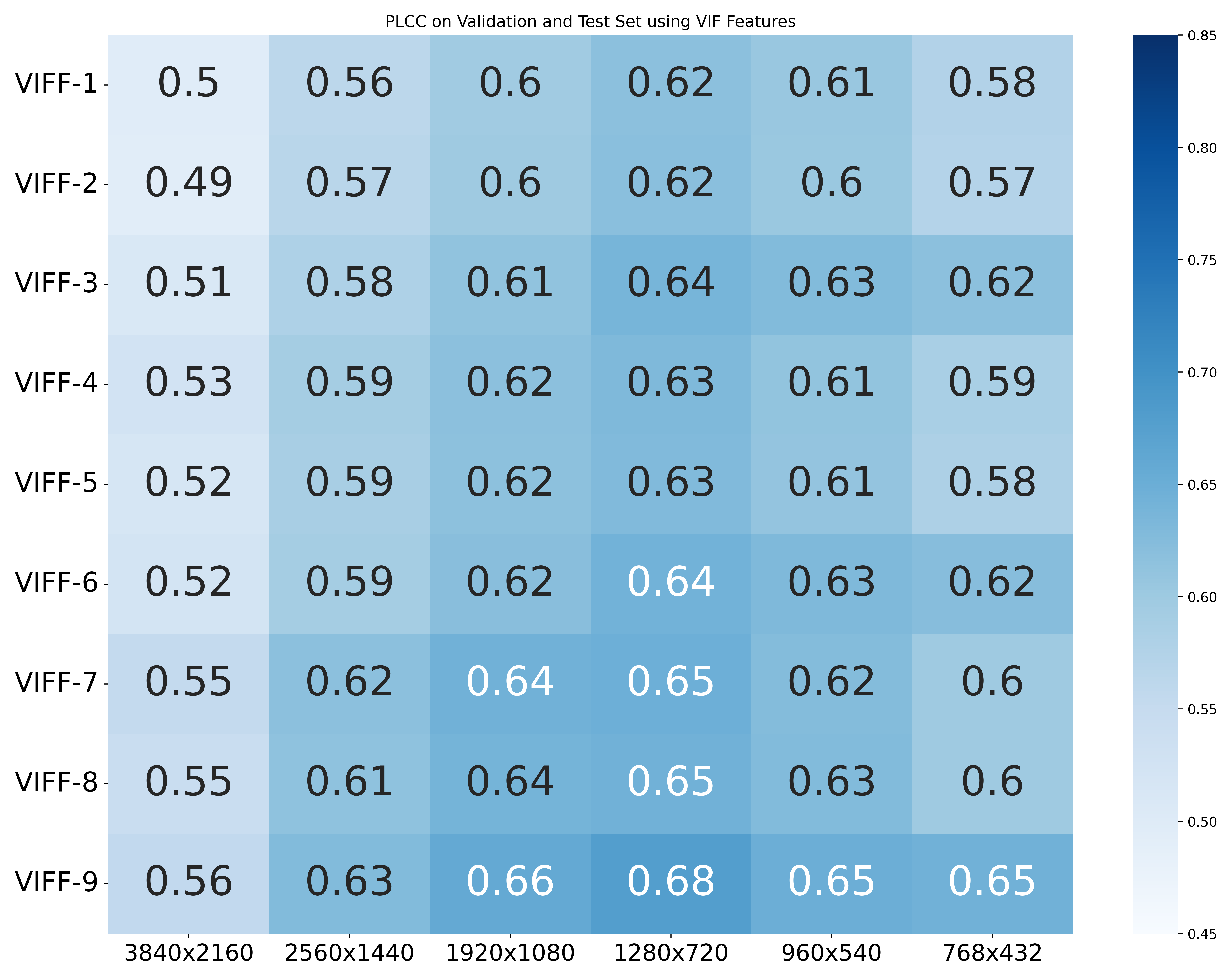}
		\caption{Quality Prediction}
	\end{subfigure}
	\begin{subfigure}[b]{0.485\linewidth}
		\centering
		\includegraphics[width=\linewidth]{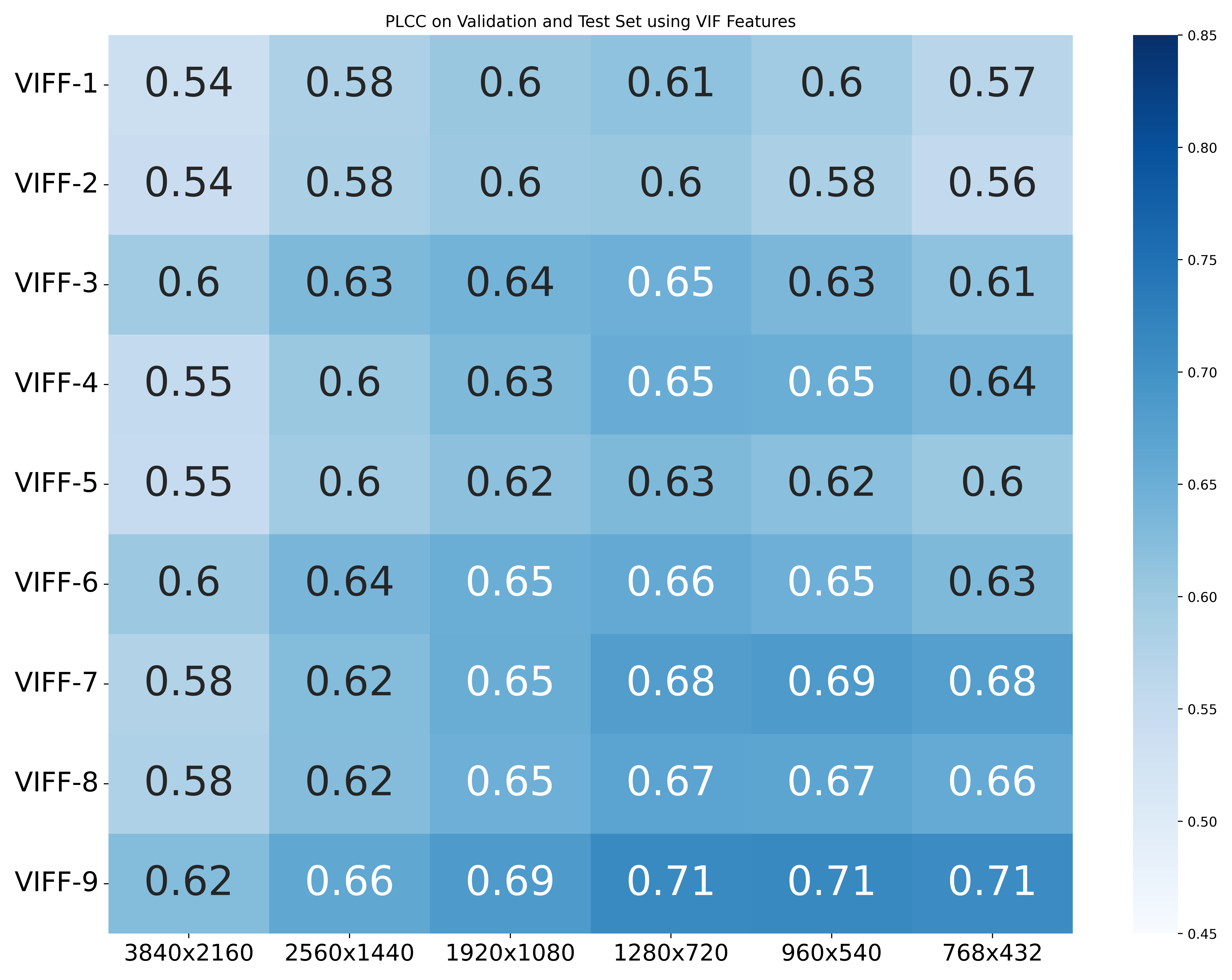}
		\caption{Bitrate Prediction}
	\end{subfigure}
	\caption{Pearson correlation coefficients between true VMAF/bitrate and VMAF/bitrate predicted using VIF feature sets over multiple resolutions on the validation and test datasets.}
	\label{fig:VIF-Results}
\end{figure}

Each VIF feature set contains features extracted from different scales and subbands of uncompressed source videos. Here we examine the performance of these models built using VIF features under different experimental settings and quality constraints. We also designed integrated models trained on VIF feature sets $\text{VIFF}_{i}$ combined with metadata like $\frac{q}{100}$, $\frac{w}{3840}$, and $\frac{h}{3840}$, to learn to predict the bitrates of compressed videos. Comparing the performances of these methods helps us to understand the accuracy, feasibility, and efficacy of constructing bitrate and quality ladders using VIF feature sets. Fig. \ref{fig:VIF-Results} shows the performances of bitrate and quality prediction models for each VIF feature set and resolution using color coding.

The models trained on VIF features extracted along each eigen vector: $\text{VIFF}_{3}$, $\text{VIFF}_{6}$, and $\text{VIFF}_{9}$, yielded better correlations against true and predicted VMAFs/bitrates, as compared to their counterparts trained on VIF features extracted along scales or subbands. The additional mean absolute luminance difference feature computed between consecutive frames improves the performance of the regressors. Similarly, regressors trained on VIF features extracted from frame differences ($\text{VIFF}_{7}$, $\text{VIFF}_{8}$, and $\text{VIFF}_{9}$) delivered superior correlations, as compared to their counterparts trained on VIF features extracted from the original frames and mean absolute luminance component differences. The models trained on $\text{VIFF}_{9}$ yielded better correlations against true and predicted VMAFs/bitrates across resolutions. Interestingly, although VIF features demonstrated higher cross-correlation (Fig. \ref{fig:cross-correlation}), the performances of the best performing models was higher than the performances of models trained on low-level features (Table \ref{table:LLF-Results}, \ref{table:LLF-Ablation-Results-Quality-Prediction} and \ref{table:LLF-Ablation-Results-Bitrate-Prediction}).

\subsection{Predicting Quality or Bitrate using an Ensemble of Low-Level features and VIF features}
Low-level features and VIF feature sets extract different and complementary features from videos. The low-level features that we use include GLCM, TC, SI, TI, CTI, CI, CF, and DCT texture energy features, while VIF feature sets are based on quality-aware Natural Scene Statistics (NSS). We considered an ensemble of low-level features and VIF features to determine whether training on their combined, complementary measurements could enhance the performance of bitrate or quality prediction models. We deployed nine different ensemble pairs from $\text{LLF}_{2}$ and $\text{VIFF}_{i}$, where $i \in \{1,2,3,...,9\}$, to combine video content dependent features along with metadata features $\log_{2}(b)$, $\frac{w}{3840}$, and $\frac{h}{3840}$, to learn video quality prediction models. We also formed nine pairs from $\text{LLF}_{3}$, $\text{VIFF}_{i}$, $i \in \{1,2,3,...,9\}$ along with $\frac{q}{100}$, $\frac{w}{3840}$, and $\frac{h}{3840}$ for which to learn bitrate prediction models.

\begin{figure}
	\centering
	\begin{subfigure}[b]{0.485\linewidth}
		\centering
		\includegraphics[width=\linewidth]{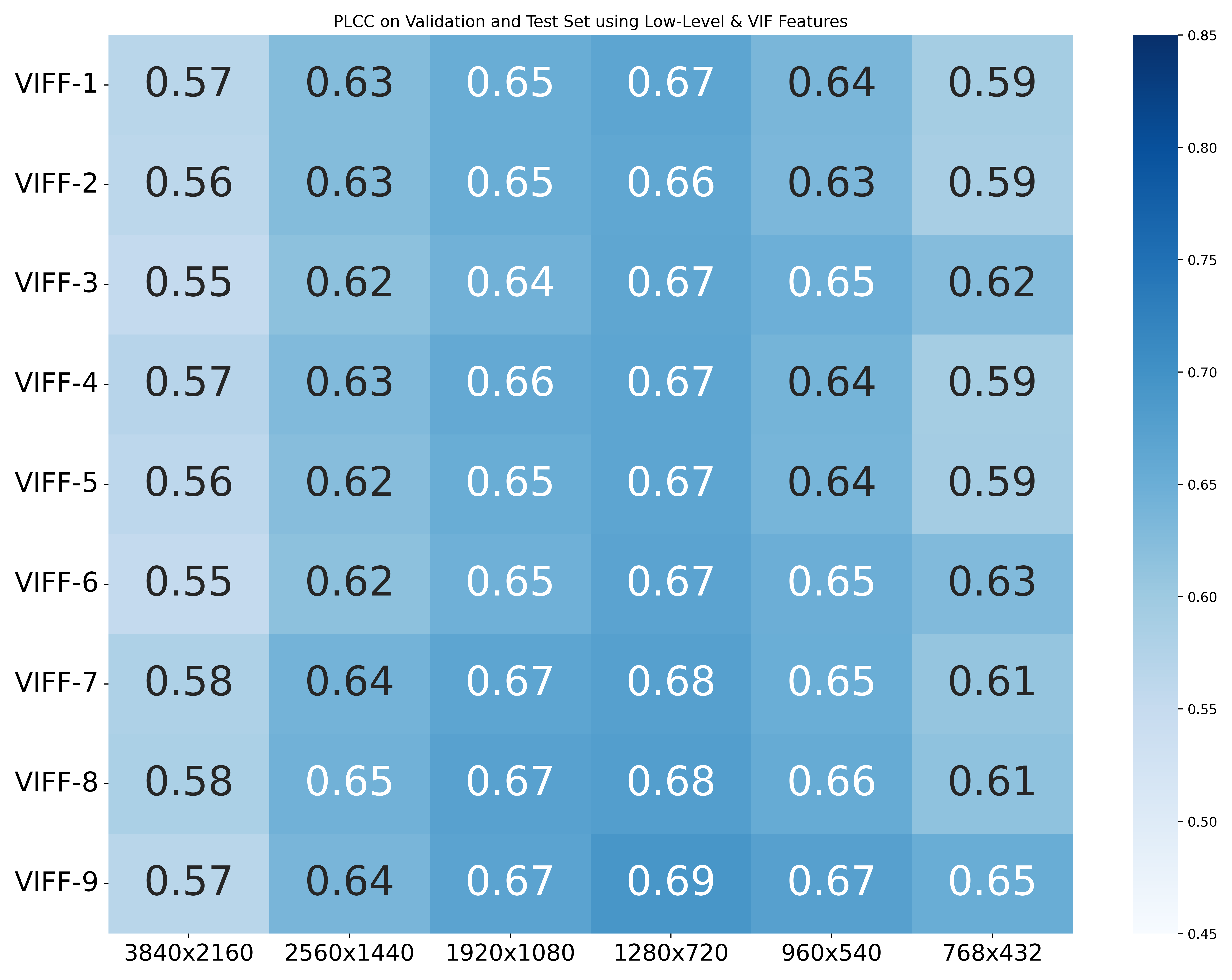}
		\caption{Quality Prediction}
	\end{subfigure}
	\begin{subfigure}[b]{0.485\linewidth}
		\centering
		\includegraphics[width=\linewidth]{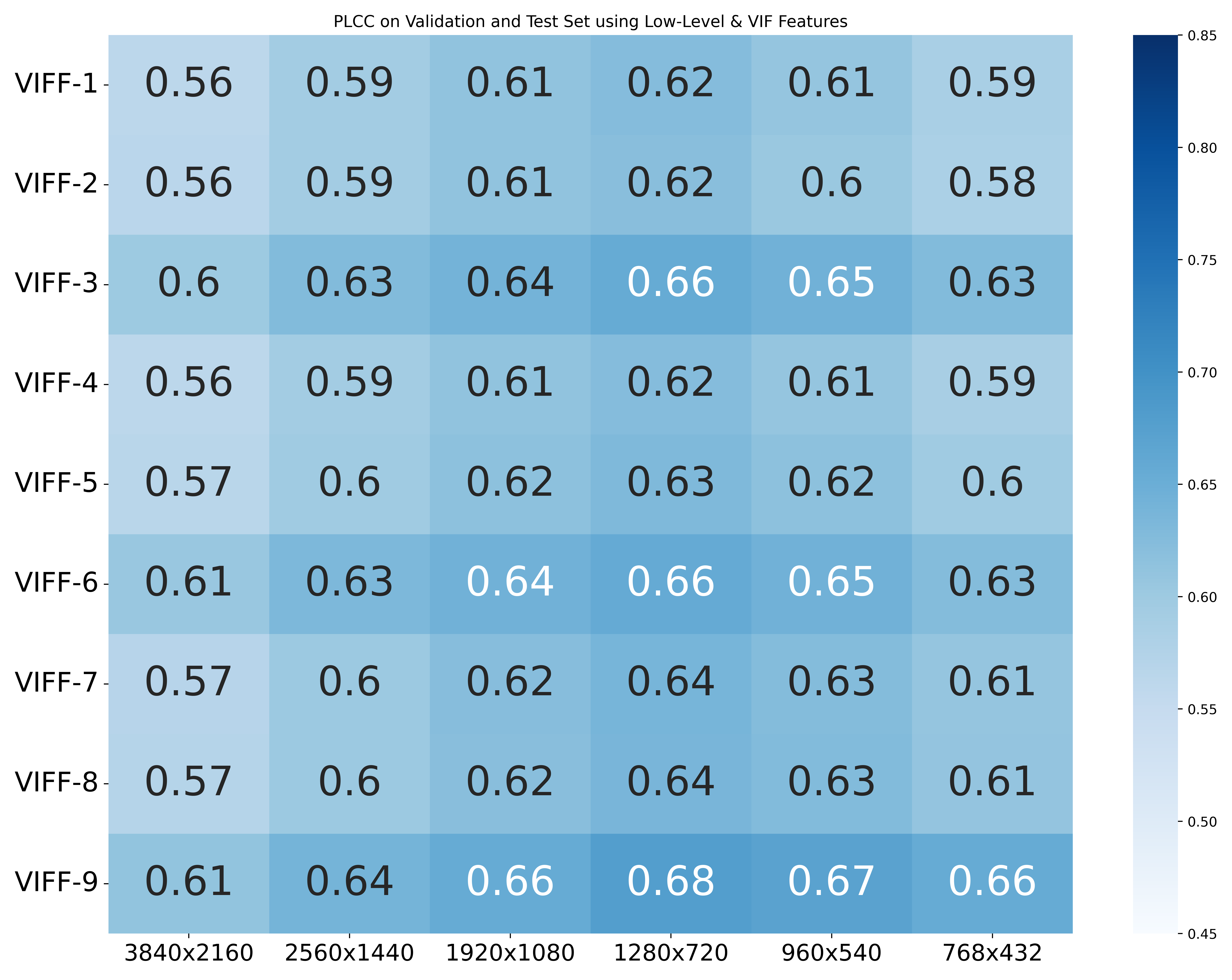}
		\caption{Bitrate Prediction}
	\end{subfigure}
	\caption{Pearson correlation coefficients between true VMAF/bitrate and VMAF/bitrate predicted using an ensemble of low-level features and VIF feature sets over multiple resolutions on the validation and test datasets.}
	\label{fig:Ensemble-LLF-VIF-Results}
\end{figure}

Fig. \ref{fig:Ensemble-LLF-VIF-Results} shows the performances of the resulting learned bitrate and quality prediction models. It may be observed that the quality prediction models were slightly more successful (better correlations) than the bitrate prediction models. The regressors trained on $\text{LLF}_{2}$/$\text{LLF}_{3}$ and $\text{VIFF}_{3}/\text{VIFF}_{6}/\text{VIFF}_{9}$ yield slightly better performance than the other models across resolutions. Similar to the performances of regressors trained on VIF features, regressors trained on low-level features and spatio-temporal VIF features delivered better performance than regressors trained on low-level features and only spatial VIF features. The performances of the best performing models was close to those models trained on either low-level features or VIF features. However, it should be understood that these correlations do not indicate the potential performance gains of these models when used to construct bitrate or quality ladders, or their BD-Rate or BD-VMAF gains, against fixed bitrate ladders.

\subsection{Correction Algorithms}
In the preceding, we have studied various techniques for predicting per-shot bitrate and quality ladders, using video quality and bitrate prediction models, respectively. Given that machine learning models are prone to errors, it is important to consider corrective measures on the predicted ladders to rectify any inaccuracies. In \cite{Bitrate-Ladder-Construction-using-Visual-Information-Fidelity}, we suggested a correction mechanism such that traversing the bitrate ladder from top to bottom (from higher to lower bitrates), will satisfy the condition that the optimal resolution at a current bitrate step should no larger than the optimal resolution at the previous step. This strategy works because:
\begin{itemize}
	\item Our dataset consists of videos compressed at six different resolutions (from 2160p to 432p) and covering a broad range of CRFs. Every resolution of each video spans a different bitrate range, with higher resolutions generally having higher quality ranges and lower resolutions having lower quality ranges.
	\item Given this, it is reasonable to assume the presence of an inherent bias within the dataset, where higher resolutions generally have higher quality ratings and vice versa. This bias acts as both an advantage and disadvantage during the learning process.
	\item During training, models learn to predict the quality of compressed videos based on their content features, bandpass statistical distributions, bitrates, and resolutions. The aforementioned bias in the dataset, which is also characteristic of rate-quality curves of videos having different resolutions and bitrates, helps these models learn to more accurately predict quality.
	\item When constructing a bitrate ladder, we use the predicted quality scores of compressed videos at each bitrate, but over all resolutions, where the resolution having the highest predicted quality is selected as the optimal resolution for encoding at that bitrate.
	\item During prediction, we have observed that the quality predictions computed on high-resolution compressed videos, tend to be higher, (particularly at lower bitrates) suggesting that the learning models overestimate quality. This is likely also due to the above mentioned bias observed in the dataset. This results in bitrate ladders where high resolutions are determined to be optimal resolutions at low bitrates.
	\item However, imposing the condition that optimal resolutions should not increase as bitrate is decreased compensates for any incorrectly predicted optimal resolutions at lower bitrates. We will refer to this as Top-to-Bottom correction.
\end{itemize}

This correction mechanism is ineffective for quality ladder prediction where we predict bitrates. During quality ladder construction, we predict bitrates across multiple resolutions at each VMAF/quality level, then the resolution associated with the lowest bitrate prediction is selected as the optimal resolution. In this case, we have observed that the dataset bias causes the predicted optimal resolutions to generally be lower than expected for generally higher quality scores. In this case, a reverse-ordering constraint: from lower to higher quality, such that the optimal resolution at a current quality step is no less than the optimal resolution at lower quality steps, compensates for the lower resolutions predicted at higher quality scores. We refer to this as Bottom-to-Top correction.

Tables \ref{table:bitrate-ladder-correction} and \ref{table:quality-ladder-correction} depict simulated examples of these Top-to-Bottom and Bottom-to-Top ladder correction strategies applied to bitrate ladders and quality ladders, respectively. These correction techniques are applied after the bitrate or quality ladders are constructed.

\begin{table}
	\caption{An example of Top-to-Bottom bitrate ladder correction.}
	\begin{subtable}{0.485\linewidth}
	\centering
	\caption{Before Correction}
	\renewcommand{\arraystretch}{1.25}
	\resizebox{\textwidth}{!}{
	\begin{tabular}{m {1.2cm} | m {1.5cm} }
		\textbf{Bitrate (in kbps)} & \textbf{Optimal Resolution} \\
		\hline
		4000 & (1920,1080) \\
		3000 & (1280,720) \\
		2000 & (1920,1080) \\
		1000 & (960,540) \\
		500 & (3840,2160) \\
	\end{tabular}}
	\end{subtable}
	\hfill
	\begin{subtable}{0.485\linewidth}
	\centering
	\caption{After Correction}
	\renewcommand{\arraystretch}{1.25}
	\resizebox{\textwidth}{!}{
	\begin{tabular}{m {1.2cm} | m {1.5cm} }
		\textbf{Bitrate (in kbps)} & \textbf{Optimal Resolution} \\
		\hline
		4000 & (1920,1080) \\
		3000 & (1280,720) \\
		2000 & (1280,720) \\
		1000 & (960,540) \\
		500 & (960,540) \\
	\end{tabular}}
	\end{subtable}
	\label{table:bitrate-ladder-correction}
\end{table}

\begin{table}
	\caption{An example of Bottom-to-Top quality ladder correction.}
	\begin{subtable}{0.485\linewidth}
	\centering
	\caption{Before Correction}
	\renewcommand{\arraystretch}{1.25}
	\resizebox{\textwidth}{!}{
	\begin{tabular}{m {1.2cm} | m {1.5cm} }
		\textbf{Quality (VMAF)} & \textbf{Optimal Resolution} \\
		\hline
		92.5 & (960,540) \\
		90 & (2560,1440) \\
		85 & (1920,1080) \\
		80 & (1280,720) \\
		75 & (1920,1080) \\
	\end{tabular}}
	\end{subtable}
	\hfill
	\begin{subtable}{0.485\linewidth}
	\centering
	\caption{After Correction}
	\renewcommand{\arraystretch}{1.25}
	\resizebox{\textwidth}{!}{
	\begin{tabular}{m {1.2cm} | m {1.5cm} }
		\textbf{Quality (VMAF)} & \textbf{Optimal Resolution} \\
		\hline
		92.5 & (2560,1440) \\
		90 & (2560,1440) \\
		85 & (1920,1080) \\
		80 & (1920,1080) \\
		75 & (1920,1080) \\
	\end{tabular}}
	\end{subtable}
	\label{table:quality-ladder-correction}
\end{table}

\subsection{Bjontegaard Delta Metrics}
One of the most commonly used techniques for evaluating or comparing the compression efficiency of video codecs, presets, or encoding modes is by calculating bitrate and quality savings using the Bjøntegaard Delta (BD) method \cite{BD}. It measures the average difference between two RQ curves by fitting a third-order cubic polynomial through the data points, and then calculating the integral to estimate the average bitrate savings, referred to as BD-Rate, and the average quality savings, referred to as BD-Quality (e.g., BD-PSNR or BD-VMAF). Although there are many implementations of this concept, based on the results in \cite{BD-Study}, we used the implementation proposed in \cite{BD-Metric}.

Since we are performing constant-quality encoding, at each step $b_{i}$ in the bitrate ladder and its corresponding optimal resolution $R_{i}$, to construct the convex hull, we utilize the RQ points of the video from the dataset having resolution of $R_{i}$, with bitrates lying within the range $[b_{i}$, $b_{i+1})$. We use this process, rather than selecting the nearest point, to obtain a precise convex hull for BD metric calculations. We applied a similar procedure when constructing convex hulls for quality ladders. We calculated the BD-Rate and BD-VMAF on each video in the validation and test datasets against the fixed bitrate ladder \cite{Fixed-Bitrate-Ladder} and the reference bitrate ladder. The reference bitrate ladders were constructed by sampling the convex hull constructed by exhaustive encoding using the bitrate ladder steps we previously mentioned. We report the mean and standard deviation of BD-Rate and BD-VMAF for each of the compared techniques, on the combined validation and test datasets.

Neither the mean nor the standard deviation of BD-Rate and BD-Quality provides sufficient information to compare the performances of the two techniques, nor help in providing any insights into whether the predicted bitrate, or the predicted quality, changes monotonically or not. So, in addition, to the mean and standard deviation of BD metrics, we calculated the closeness of the performance of a method to the exhaustive reference bitrate ladder, by estimating the fractions of samples yielding both BD-Rate savings and BD-VMAF gain greater than 75\%, 50\%, and 25\% of BD metrics of reference bitrate ladder against fixed bitrate ladder, which we denote by $\text{f}_{75}$, $\text{f}_{50}$, and $\text{f}_{25}$, respectively. We also performed a monotonicity check to determine that the predicted bitrates or qualities monotonically change at each resolution of every video file in the test and validation datasets. This also helps to determine whether the learned models overfit on the training dataset, \textit{viz.}, learned a non-monotonic rate quality mapping. We trained the bitrate and quality prediction models with hyperparameters such that there were no regular failures of monotonicity.

\subsection{Constructing Per-Shot Bitrate and Quality Ladders}
\begin{table*}
	\normalfont
	\normalsfcodes
	\renewcommand{\arraystretch}{1.75}
	\centering
	\caption{Means and standard deviations of BD-metrics, and closeness of each model's predicted per-shot bitrate ladders against fixed and reference bitrate ladders on the validation and test datasets. Bitrate ladders have the following bitrates (in kbps) as steps: [500, 1000, 2000, 3000, 4000, 5000, 6000, 7000, 8000, 9000, 10500, 12000, 15000].}
	\resizebox{\textwidth}{!}{
	\begin{tabular}{| m{4.25cm} | c | c | c | c | c | c | c |}

	\hline
	\textbf{Features Set} & \multicolumn{2}{ c |}{\textbf{BL vs Fixed Bitrate Ladder}} & \multicolumn{2}{ c |}{\textbf{BL vs Reference Bitrate Ladder}} & $\textbf{f}_{25}$ & $\textbf{f}_{50}$ & $\textbf{f}_{75}$ \\
	\cline{2-5}
	& BD-Rate (in \%) & BD-VMAF & BD-Rate (in \%) & BD-VMAF & & &\\
	\hline

	$\log_{2}(b)$, $\frac{w}{3840}$, $\frac{h}{3840}$ & $-9.256/20.964$ & $2.084/4.369$ & $14.053/11.451$ & $-2.144/1.933$ & $0.500$ & $0.433$ & $0.167$\\

	\hline

	$\text{LLF}_{2}$, $\log_{2}(b)$, $\frac{w}{3840}$, $\frac{h}{3840}$ & $-20.429/16.747$ & $4.323/4.092$ & $0.439/4.791$ & $-0.016/1.004$ & $0.967$ & $0.933$ & $0.767$\\

	\hline

	$\text{VIFF}_{1}$, $\log_{2}(b)$, $\frac{w}{3840}$, $\frac{h}{3840}$ & $-12.238/22.332$ & $2.790/4.704$ & $9.780/13.017$ & $-1.519/2.145$ & $0.700$ & $0.600$ & $0.400$\\

	$\text{VIFF}_{2}$, $\log_{2}(b)$, $\frac{w}{3840}$, $\frac{h}{3840}$ & $-16.022/20.536$ & $3.521/4.678$ & $5.381/8.955$ & $-0.922/1.617$ & $0.767$ & $0.767$ & $0.633$\\

	$\text{VIFF}_{3}$, $\log_{2}(b)$, $\frac{w}{3840}$, $\frac{h}{3840}$ & $-17.372/18.170$ & $3.665/4.184$ & $3.746/5.768$ & $-0.628/1.039$ & $0.767$ & $0.700$ & $0.567$\\

	$\text{VIFF}_{4}$, $\log_{2}(b)$, $\frac{w}{3840}$, $\frac{h}{3840}$ & $-16.852/18.941$ & $3.624/4.184$ & $4.613/7.811$ & $-0.717/1.253$ & $0.767$ & $0.700$ & $0.567$\\

	$\text{VIFF}_{5}$, $\log_{2}(b)$, $\frac{w}{3840}$, $\frac{h}{3840}$ & $-17.822/18.324$ & $3.853/4.243$ & $3.396/7.565$ & $-0.463/1.213$ & $0.800$ & $0.767$ & $0.633$\\

	$\text{VIFF}_{6}$, $\log_{2}(b)$, $\frac{w}{3840}$, $\frac{h}{3840}$ & $-17.655/18.778$ & $3.756/4.275$ & $3.661/6.338$ & $-0.628/1.089$ & $0.767$ & $0.700$ & $0.600$\\

	$\text{VIFF}_{7}$, $\log_{2}(b)$, $\frac{w}{3840}$, $\frac{h}{3840}$ & $-16.596/17.158$ & $3.618/4.013$ & $5.186/8.562$ & $-0.659/1.278$ & $0.833$ & $0.767$ & $0.567$\\

	$\text{VIFF}_{8}$, $\log_{2}(b)$, $\frac{w}{3840}$, $\frac{h}{3840}$ & $-17.308/17.842$ & $3.714/4.174$ & $4.250/7.433$ & $-0.637/1.289$ & $0.800$ & $0.700$ & $0.567$\\

	$\text{VIFF}_{9}$, $\log_{2}(b)$, $\frac{w}{3840}$, $\frac{h}{3840}$ & $-17.764/17.275$ & $3.689/3.916$ & $3.529/6.726$ & $-0.525/1.164$ & $0.800$ & $0.767$ & $0.600$\\

	\hline

	$\text{LLF}_{2}$, $\text{VIFF}_{1}$, $\log_{2}(b)$, $\frac{w}{3840}$, $\frac{h}{3840}$ & $-20.151/16.703$ & $4.195/3.889$ & $1.025/4.703$ & $-0.086/0.933$ & $0.967$ & $0.867$ & $0.733$\\

	$\text{LLF}_{2}$, $\text{VIFF}_{2}$, $\log_{2}(b)$, $\frac{w}{3840}$, $\frac{h}{3840}$ & $-20.523/17.207$ & $4.390/4.071$ & $0.829/5.388$ & $-0.065/1.073$ & $0.967$ & $0.867$ & $0.767$\\

	$\text{LLF}_{2}$, $\text{VIFF}_{3}$, $\log_{2}(b)$, $\frac{w}{3840}$, $\frac{h}{3840}$ & $-20.201/16.083$ & $4.259/3.850$ & $1.113/5.015$ & $-0.089/0.943$ & $1.000$ & $0.900$ & $0.800$\\

	$\text{LLF}_{2}$, $\text{VIFF}_{4}$, $\log_{2}(b)$, $\frac{w}{3840}$, $\frac{h}{3840}$ & $-20.314/16.231$ & $4.327/3.981$ & $1.005/5.852$ & $-0.051/0.987$ & $0.967$ & $0.933$ & $0.833$\\

	$\text{LLF}_{2}$, $\text{VIFF}_{5}$, $\log_{2}(b)$, $\frac{w}{3840}$, $\frac{h}{3840}$ & $-20.379/16.601$ & $4.283/3.864$ & $0.764/4.318$ & $-0.047/0.846$ & $0.933$ & $0.867$ & $0.800$\\

	$\text{LLF}_{2}$, $\text{VIFF}_{6}$, $\log_{2}(b)$, $\frac{w}{3840}$, $\frac{h}{3840}$ & $-\textbf{20.688}/16.624$ & $\textbf{4.425}/4.032$ & $\textbf{0.347}/4.544$ & $\textbf{0.010}/0.916$ & $0.967$ & $0.933$ & $0.833$\\

	$\text{LLF}_{2}$, $\text{VIFF}_{7}$, $\log_{2}(b)$, $\frac{w}{3840}$, $\frac{h}{3840}$ & $-20.473/16.738$ & $4.288/3.943$ & $0.824/4.721$ & $-0.080/0.979$ & $0.967$ & $0.867$ & $0.667$\\

	$\text{LLF}_{2}$, $\text{VIFF}_{8}$, $\log_{2}(b)$, $\frac{w}{3840}$, $\frac{h}{3840}$ & $-20.373/17.039$ & $4.374/4.016$ & $0.794/3.809$ & $-0.067/0.756$ & $0.967$ & $0.833$ & $0.767$\\

	$\text{LLF}_{2}$, $\text{VIFF}_{9}$, $\log_{2}(b)$, $\frac{w}{3840}$, $\frac{h}{3840}$ & $-20.086/16.797$ & $4.267/4.095$ & $1.037/5.256$ & $-0.092/1.008$ & $0.967$ & $0.800$ & $0.733$\\

	\hline

	$\text{LLF}_{1}$ (Cross-Over bitrates) & $-20.430/15.853$ & $4.361/3.702$ & $1.342/5.154$ & $-0.124/0.840$ & $\mathbf{0.967}$ & $\mathbf{0.933}$ & $\mathbf{0.900}$\\

	\hline
	\end{tabular}}
	\label{table:bitrate-ladder-bd-metrics}
\end{table*}

The construction of per-shot reference bitrate ladders is achieved by sampling the convex hulls computed on videos from both the validation and test datasets. The reference bitrate ladder demonstrated a BD-Rate performance having a mean improvement of $-20.63\%$ (negative BD-rate means a gain in bitrate) and a standard deviation of $17.124\%$, and a BD-VMAF performance having a mean improvement of $4.473$ (positive means gain in quality) and standard deviation $4.139$ against the fixed bitrate ladder \cite{Fixed-Bitrate-Ladder}. Our models were compared against techniques that predict cross-over bitrates or cross-over VMAF points. For more accurate comparisons, we computed the performance of bitrate and quality ladders constructed using true cross-over bitrates and VMAFs points, respectively. We found that the bitrate ladder constructed using true cross-over bitrates, attained a BD-rate performance having a mean improvement of $-21.653\%$, and a standard deviation of $16.835\%$. It also obtained a BD-VMAF performance having a mean improvement of $4.712$ and a standard deviation of $4.091$. In a similar vein, the quality ladder constructed using true cross-over VMAF points, demonstrated a BD-rate performance having a mean improvement of $-21.847\%$ and a standard deviation of $16.254\%$, and a BD-VMAF performance having a mean improvement of $4.752$ and standard deviation $4.145$.

\begin{table*}
	\normalfont
	\normalsfcodes
	\renewcommand{\arraystretch}{1.75}
	\centering
	\caption{Means and standard deviations of BD-metrics, and closeness of each model's predicted per-shot quality ladders against fixed and reference bitrate ladders on the validation and test datasets. Quality ladders have the following VMAF scores as steps: [25, 35, 45, 50, 55, 60, 65, 70, 75, 80, 85, 90, 92.5].}
	\resizebox{\textwidth}{!}{
	\begin{tabular}{| m{4.25cm} | c | c | c | c | c | c | c |}

	\hline
	\textbf{Features Set} & \multicolumn{2}{ c |}{\textbf{QL vs Fixed Bitrate Ladder}} & \multicolumn{2}{ c |}{\textbf{QL vs Reference Bitrate Ladder}} & $\textbf{f}_{25}$ & $\textbf{f}_{50}$ & $\textbf{f}_{75}$ \\
	\cline{2-5}
	& BD-Rate (in \%) & BD-VMAF & BD-Rate (in \%) & BD-VMAF & & & \\
	\hline

	$q/100$, $\frac{w}{3840}$, $\frac{h}{3840}$ & $-15.311/15.217$ & $3.042/3.383$ & $12.546/14.353$ & $-2.422/2.534$ & $0.767$ & $0.667$ & $0.533$\\

	\hline

	$\text{LLF}_{3}$, $q/100$, $\frac{w}{3840}$, $\frac{h}{3840}$ & $-20.546/16.722$ & $\textbf{4.599}/4.107$ & $0.615/3.546$ & $-0.050/0.786$ & $1.000$ & $0.900$ & $0.900$\\

	\hline

	$\text{VIFF}_{1}$, $q/100$, $\frac{w}{3840}$, $\frac{h}{3840}$ & $-18.337/17.154$ & $4.064/4.225$ & $3.274/5.763$ & $-0.513/1.069$ & $0.833$ & $0.767$ & $0.600$\\

	$\text{VIFF}_{2}$, $q/100$, $\frac{w}{3840}$, $\frac{h}{3840}$ & $-17.300/17.353$ & $3.728/4.149$ & $4.524/5.619$ & $-0.747/1.024$ & $0.833$ & $0.733$ & $0.533$\\

	$\text{VIFF}_{3}$, $q/100$, $\frac{w}{3840}$, $\frac{h}{3840}$ & $-17.905/16.652$ & $3.905/4.149$ & $3.829/4.074$ & $-0.597/0.715$ & $0.867$ & $0.800$ & $0.633$\\

	$\text{VIFF}_{4}$, $q/100$, $\frac{w}{3840}$, $\frac{h}{3840}$ & $-18.344/16.510$ & $4.201/4.197$ & $3.070/5.642$ & $-0.425/0.963$ & $0.900$ & $0.867$ & $0.667$\\

	$\text{VIFF}_{5}$, $q/100$, $\frac{w}{3840}$, $\frac{h}{3840}$ & $-17.584/17.344$ & $3.880/4.229$ & $4.106/5.560$ & $-0.637/1.051$ & $0.733$ & $0.700$ & $0.533$\\

	$\text{VIFF}_{6}$, $q/100$, $\frac{w}{3840}$, $\frac{h}{3840}$ & $-18.932/16.229$ & $4.265/4.124$ & $2.724/4.890$ & $-0.451/1.031$ & $0.933$ & $0.867$ & $0.700$\\

	$\text{VIFF}_{7}$, $q/100$, $\frac{w}{3840}$, $\frac{h}{3840}$ & $-17.426/17.025$ & $3.847/4.138$ & $4.456/6.497$ & $-0.652/1.159$ & $0.800$ & $0.800$ & $0.633$\\

	$\text{VIFF}_{8}$, $q/100$, $\frac{w}{3840}$, $\frac{h}{3840}$ & $-17.418/16.493$ & $3.800/4.033$ & $4.692/5.700$ & $-0.722/1.008$ & $0.833$ & $0.767$ & $0.533$\\

	$\text{VIFF}_{9}$, $q/100$, $\frac{w}{3840}$, $\frac{h}{3840}$ & $-18.452/16.350$ & $4.104/4.046$ & $3.031/4.573$ & $-0.435/0.866$ & $0.900$ & $0.800$ & $0.733$\\

	\hline

	$\text{LLF}_{3}$, $\text{VIFF}_{1}$, $q/100$, $\frac{w}{3840}$, $\frac{h}{3840}$ & $-20.514/16.740$ & $4.549/4.054$ & $0.471/3.690$ & $-0.015/0.822$ & $0.933$ & $0.933$ & $0.867$\\

	$\text{LLF}_{3}$, $\text{VIFF}_{2}$, $q/100$, $\frac{w}{3840}$, $\frac{h}{3840}$ & $-20.138/16.605$ & $4.437/3.985$ & $0.940/3.834$ & $-0.103/0.847$ & $1.000$ & $0.900$ & $0.833$\\

	$\text{LLF}_{3}$, $\text{VIFF}_{3}$, $q/100$, $\frac{w}{3840}$, $\frac{h}{3840}$ & $-20.500/16.748$ & $4.489/4.041$ & $0.545/3.815$ & $-0.022/0.832$ & $0.967$ & $0.933$ & $0.833$\\

	$\text{LLF}_{3}$, $\text{VIFF}_{4}$, $q/100$, $\frac{w}{3840}$, $\frac{h}{3840}$ & $-20.727/16.557$ & $4.560/4.110$ & $0.294/3.503$ & $0.028/0.768$ & $0.967$ & $0.933$ & $0.867$\\

	$\text{LLF}_{3}$, $\text{VIFF}_{5}$, $q/100$, $\frac{w}{3840}$, $\frac{h}{3840}$ & $-20.690/16.663$ & $4.568/4.005$ & $0.329/3.495$ & $-0.000/0.793$ & $0.967$ & $0.933$ & $0.867$\\

	$\text{LLF}_{3}$, $\text{VIFF}_{6}$, $q/100$, $\frac{w}{3840}$, $\frac{h}{3840}$ & $-20.686/16.501$ & $4.533/3.976$ & $0.438/3.477$ & $-0.004/0.780$ & $0.967$ & $0.967$ & $0.933$\\

	$\text{LLF}_{3}$, $\text{VIFF}_{7}$, $q/100$, $\frac{w}{3840}$, $\frac{h}{3840}$ & $-20.396/16.504$ & $4.506/4.122$ & $0.610/3.592$ & $-0.030/0.789$ & $0.967$ & $0.900$ & $0.867$\\

	$\text{LLF}_{3}$, $\text{VIFF}_{8}$, $q/100$, $\frac{w}{3840}$, $\frac{h}{3840}$ & $-20.680/16.617$ & $4.551/4.009$ & $0.310/3.503$ & $0.013/0.786$ & $\textbf{1.000}$ & $\textbf{0.967}$ & $\textbf{0.900}$\\

	$\text{LLF}_{3}$, $\text{VIFF}_{9}$, $q/100$, $\frac{w}{3840}$, $\frac{h}{3840}$ & $-20.538/16.449$ & $4.530/3.972$ & $0.570/3.638$ & $-0.030/0.818$ & $0.967$ & $0.967$ & $0.867$\\

	\hline

	$\text{LLF}_{1}$ (Cross-Over VMAFs) & $-\textbf{20.863}/16.728$ & $4.545/4.247$ & $\mathbf{-0.129}/3.304$ & $\mathbf{0.102}/0.741$ & $0.933$ & $0.967$ & $0.867$\\

	\hline
	\end{tabular}}
	\label{table:quality-ladder-bd-metrics}
\end{table*}

The performances of the per-shot bitrate ladders and quality ladders, constructed using the various described methods, are presented in Tables \ref{table:bitrate-ladder-bd-metrics} and \ref{table:quality-ladder-bd-metrics}, respectively. These are compared against a fixed bitrate ladder \cite{Fixed-Bitrate-Ladder} and reference exhaustive bitrate ladders. In Table \ref{table:bitrate-ladder-bd-metrics}, all rows, except the last one, show the performances of bitrate ladders constructed using quality prediction methods. The last row presents the performance of bitrate ladders constructed by predicting cross-over bitrates. Similarly, in Table \ref{table:quality-ladder-bd-metrics}, all rows, barring the last one, list the performances of quality ladders constructed using bitrate prediction methods. The last row of the Table shows the performance of quality ladders constructed by predicting cross-over VMAF points. The last three columns of the Tables demonstrate the similarity of the predicted per-shot bitrate and quality ladders with the reference bitrate ladder, when comparing their performance relative to the fixed bitrate ladder. The metrics $\text{f}_{75}$, $\text{f}_{50}$, and $\text{f}_{25}$ provide insights into the relative performances of bitrate and quality ladders, including the reference bitrate ladder. For each method, a high value of $\text{f}_{i}$ signifies that a large proportion of test samples, encompassing both validation and test datasets, exhibit a BD-Rate and BD-VMAF gains more than $i\%$ of the BD-Rate and BD-VMAF gains of the reference bitrate ladder against the fixed bitrate ladder \cite{Fixed-Bitrate-Ladder}.

From Table \ref{table:bitrate-ladder-bd-metrics} and Table \ref{table:quality-ladder-bd-metrics}, it may be observed that the per-shot bitrate and quality ladders constructed by predicting cross-over bitrates and VMAFs using the low-level features $\text{LLF}_{1}$ demonstrated excellent performance in BD-rate and BD-VMAF as compared to the fixed bitrate ladder \cite{Fixed-Bitrate-Ladder} and the reference bitrate ladder. The per-shot bitrate ladders constructed using cross-over bitrates demonstrated a mean BD-rate loss of $1.342\%$ and a mean BD-VMAF loss of $-0.124$ versus the reference bitrate ladder. Similarly, the per-shot quality ladders showed a mean BD-rate loss of $-0.129\%$ and a mean BD-VMAF gain of $0.102$ against the reference bitrate ladder. The negative value BD-rate against the reference bitrate ladder was likely due to noise introduced by sampling from the convex hull. Their performance is close to that of ladders constructed using true cross-over points.

From Table \ref{table:bitrate-ladder-bd-metrics} and Table \ref{table:quality-ladder-bd-metrics}, it is evident that per-shot bitrate and quality ladders constructed using content-dependent video features extracted from the source video demonstrated improved mean BD-Rate and BD-VMAF performance relative to the fixed bitrate ladder \cite{Fixed-Bitrate-Ladder} and the reference bitrate ladder. This is evident when the performances of our method is compared to that of ladders constructed using regressors trained on metadata. These results demonstrate the effectiveness of content-adaptive features, while also establishing a baseline against which the efficacy of video features can be determined. Regressors trained on the VIF features sets yielded significant gains in terms of both quality and bitrates, as compared to fixed bitrate ladders \cite{Fixed-Bitrate-Ladder}. Similar to the high correlations observed between true and predicted VMAF/bitrates, in most cases regressors trained on VIF features extracted along eigen-vectors demonstrated better mean BD-Rate and BD-VMAF performance than regressors trained on features from scales and subbands. The addition of temporal features such as mean average luminance differences, and VIF features extracted on frame differences, also improved the accuracies of predicted bitrate and quality ladders. Although their performances against the reference bitrate ladder is not excellent, bitrate/quality ladders constructed using regressors trained on VIF features showed more than 50\% gains of the reference bitrate ladder on approximately 70\% of samples in the test dataset. The regressors trained on the $\text{VIFF}_{9}$ features demonstrated the best performance in terms of BD-metrics against the fixed bitrate ladder and the reference bitrate ladder.

Although, they achieved slightly lower correlations than did regressors trained on VIF features, the performances of per-shot ladders constructed using regressors trained on low-level features yielded better mean BD-Rate and BD-VMAF performance against the fixed bitrate ladder and the reference bitrate ladders. While the correlation gains obtained against the true VMAF/bitrates were significant, it is worth noting that the actual performances of bitrate/quality ladders is determined by the relative ordering of the predicted quality/bitrates at each resolution. Consequently, the representation of video content provided by the low-level features likely contributed to improved performance of bitrate/quality ladders constructed using these low-level features. Tables \ref{table:bitrate-ladder-bd-metrics-LLF} and \ref{table:quality-ladder-bd-metrics-LLF} show the performances of per-shot bitrate and quality ladders constructed using subsets of the low-level features. The mean BD-rate losses against the reference bitrate ladders demonstrated by subsets of the low-level features was about 2\% lower than the losses delivered using best performing VIF feature set ($\text{VIFF}_{9}$). Further, the performances of ladders constructed using the $\text{LLF}_{2}$ and $\text{LLF}_{3}$ features were much better than the performances of ladders constructed using subsets of the low-level features.

The regressors that were trained on the ensemble of low-level features and VIF features demonstrated better performance than either using only low-level features or VIF features. It may be observed that these methods produce mean BD-rate gains of at least 20\% and BD-VMAF gains of at least 4 against the fixed bitrate ladder. Based on closeness, the per-shot bitrate ladders constructed using the ensemble of these features were only similar or slightly better than the per-shot bitrate ladders constructed using low-level features. Interestingly, the mean BD-rate and BD-VMAF values were slightly better for the per-shot bitrate ladders constructed using low-level features, on most of the feature sets. In most cases, the per-shot quality ladders constructed using the ensemble of low-level features and VIF features yielded outcomes that were closer to the reference bitrate ladder, with better mean BD-rate and BD-VMAF values, than those constructed using low-level features. These results show the effectiveness of using the ensemble of low-level features and VIF features, when constructing per-shot bitrate and quality ladders. Based on mean BD-rate and BD-VMAF values and closeness, the per-shot bitrate and quality ladders constructed using low-level features $\text{LLF}_{2}$, $\text{VIFF}_{6}$, and $\text{LLF}_{3}$, $\text{VIFF}_{8}$, respectively, delivered the best performance.

\begin{figure*}
	\centering
	\begin{subfigure}[b]{0.245\linewidth}
	\centering
	\includegraphics[width=\linewidth]{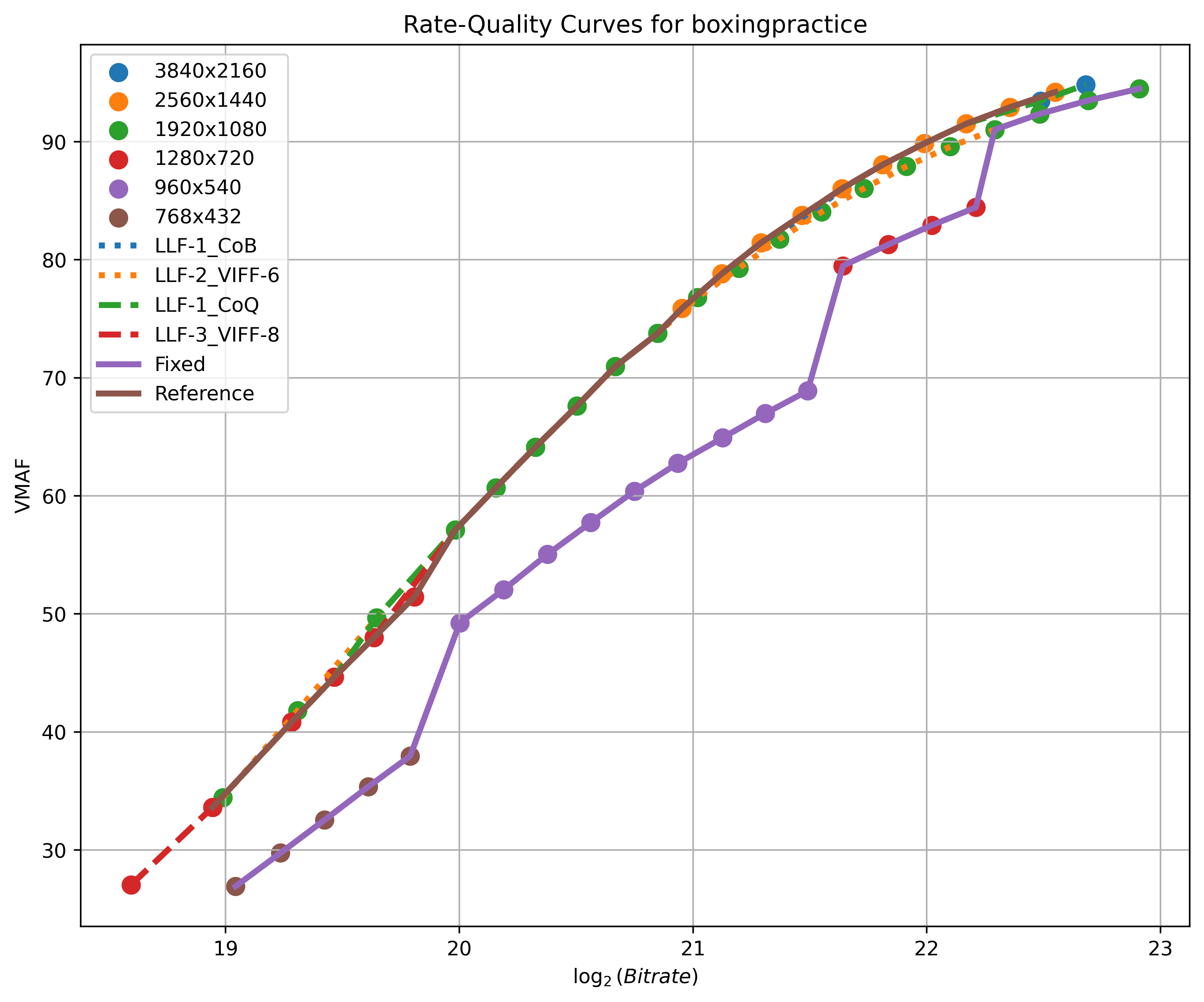}
	\end{subfigure}
	\begin{subfigure}[b]{0.245\linewidth}
	\centering
	\includegraphics[width=\linewidth]{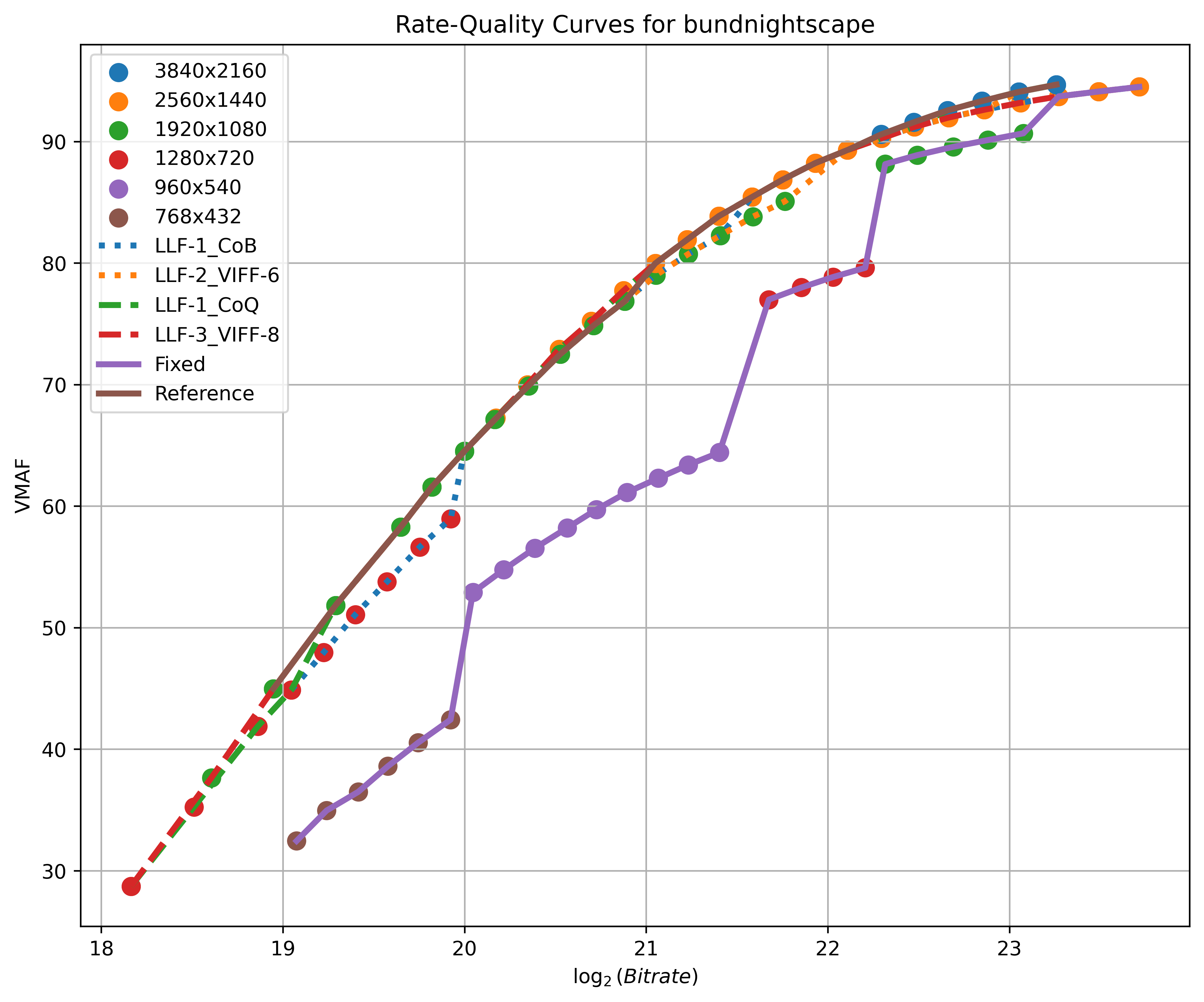}
	\end{subfigure}
	\begin{subfigure}[b]{0.245\linewidth}
	\centering
	\includegraphics[width=\linewidth]{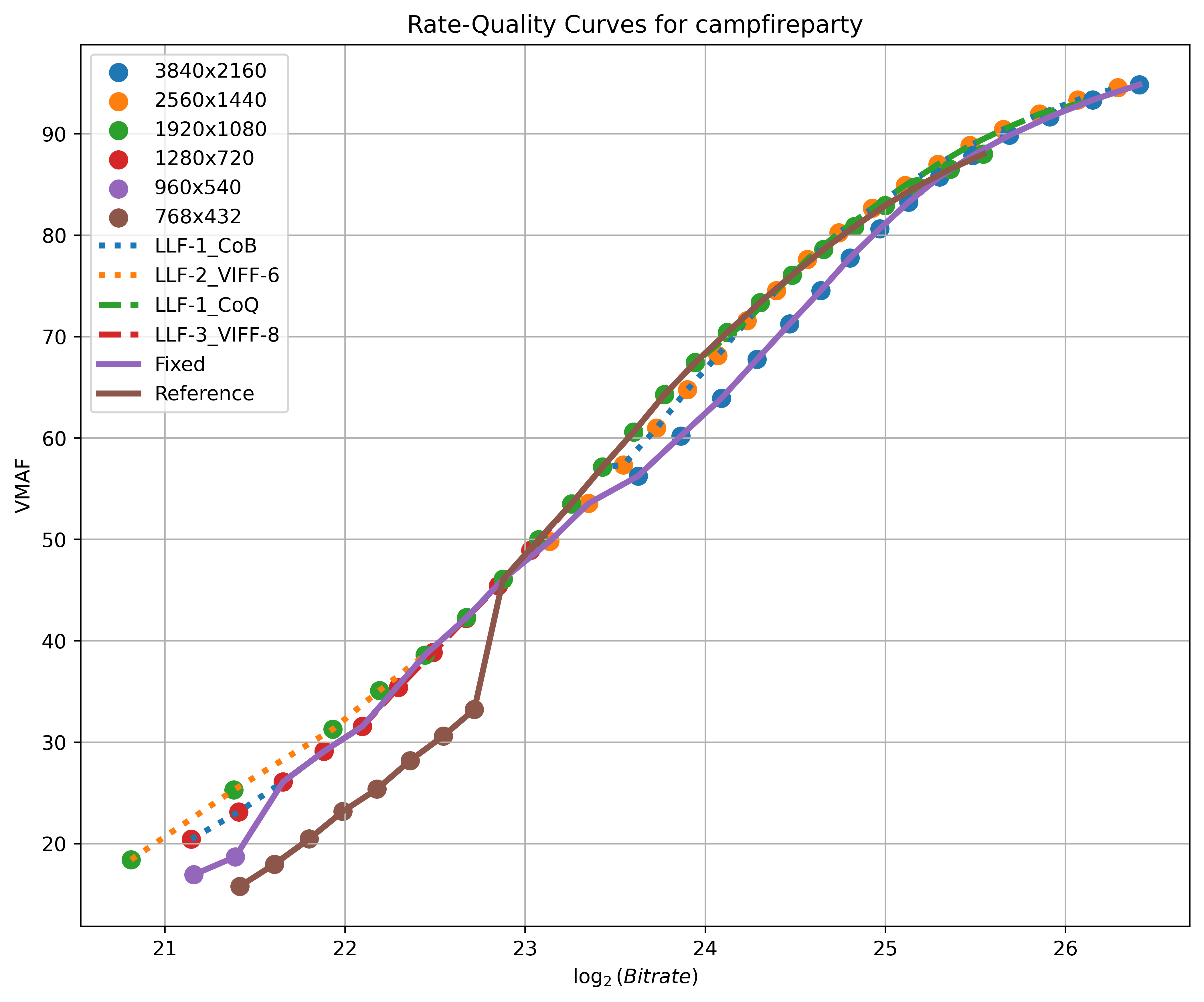}
	\end{subfigure}
	\begin{subfigure}[b]{0.245\linewidth}
	\centering
	\includegraphics[width=\linewidth]{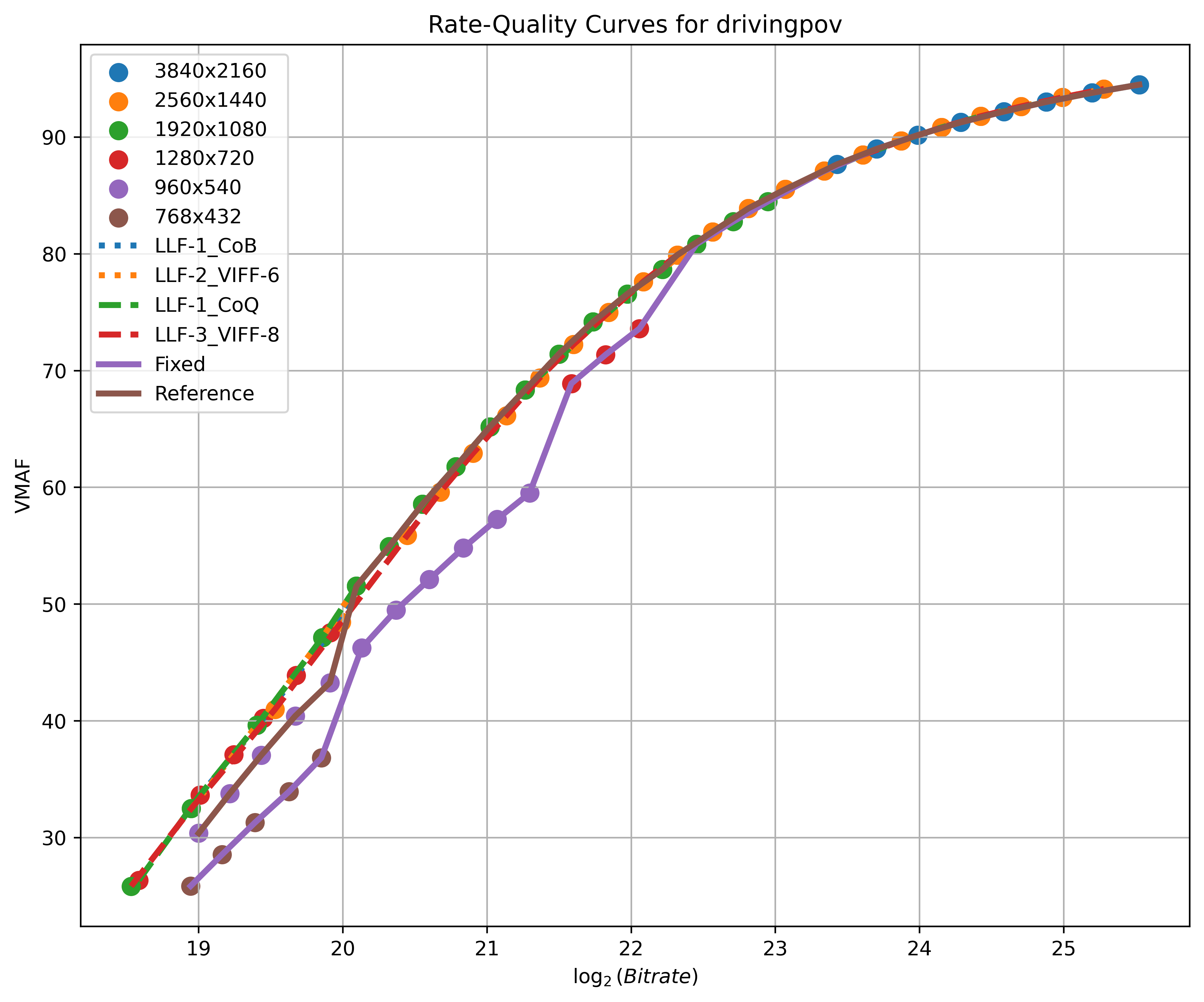}
	\end{subfigure}
	\begin{subfigure}[b]{0.245\linewidth}
	\centering
	\includegraphics[width=\linewidth]{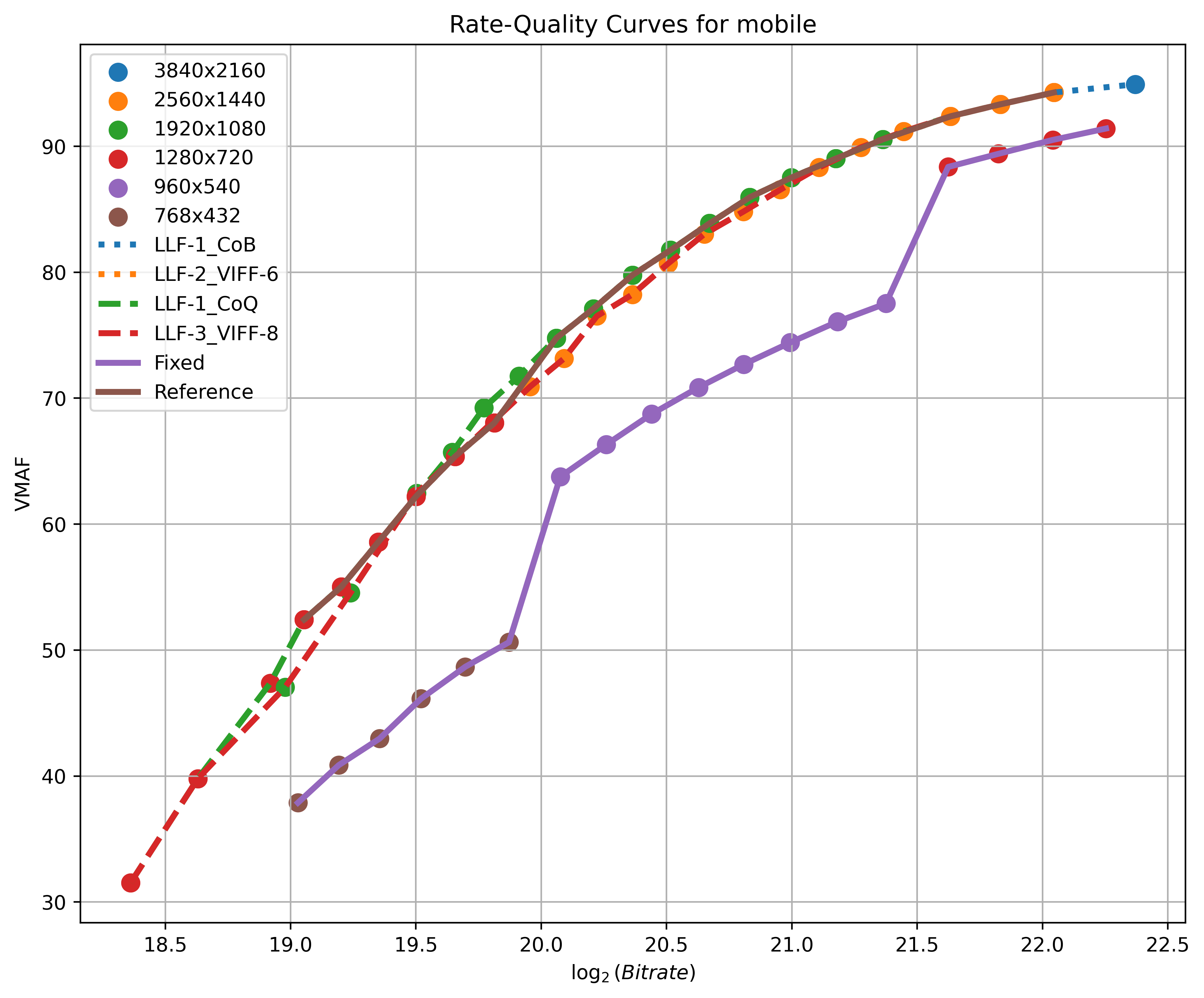}
	\end{subfigure}
	\begin{subfigure}[b]{0.245\linewidth}
	\centering
	\includegraphics[width=\linewidth]{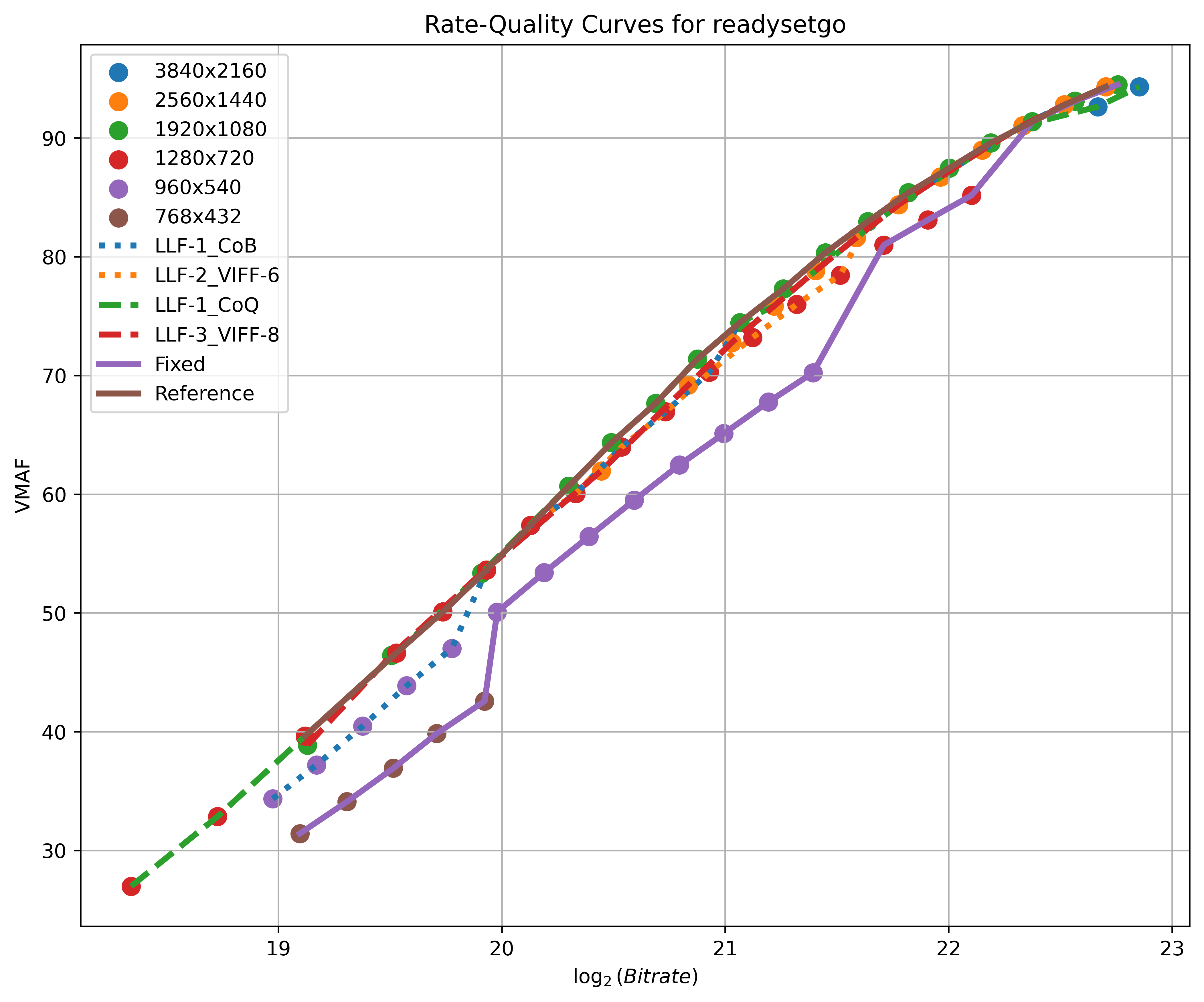}
	\end{subfigure}
	\begin{subfigure}[b]{0.245\linewidth}
	\centering
	\includegraphics[width=\linewidth]{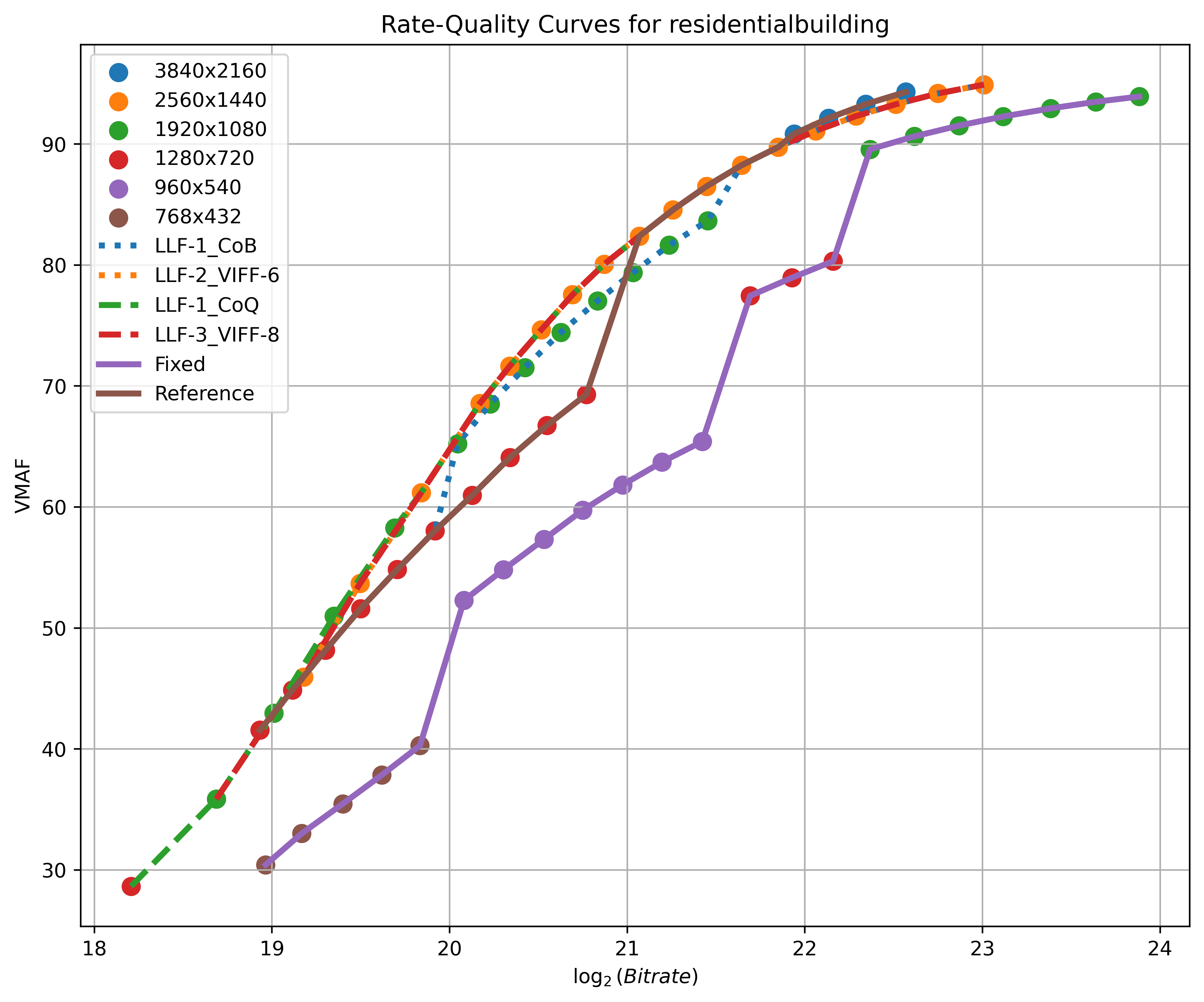}
	\end{subfigure}
	\begin{subfigure}[b]{0.245\linewidth}
	\centering
	\includegraphics[width=\linewidth]{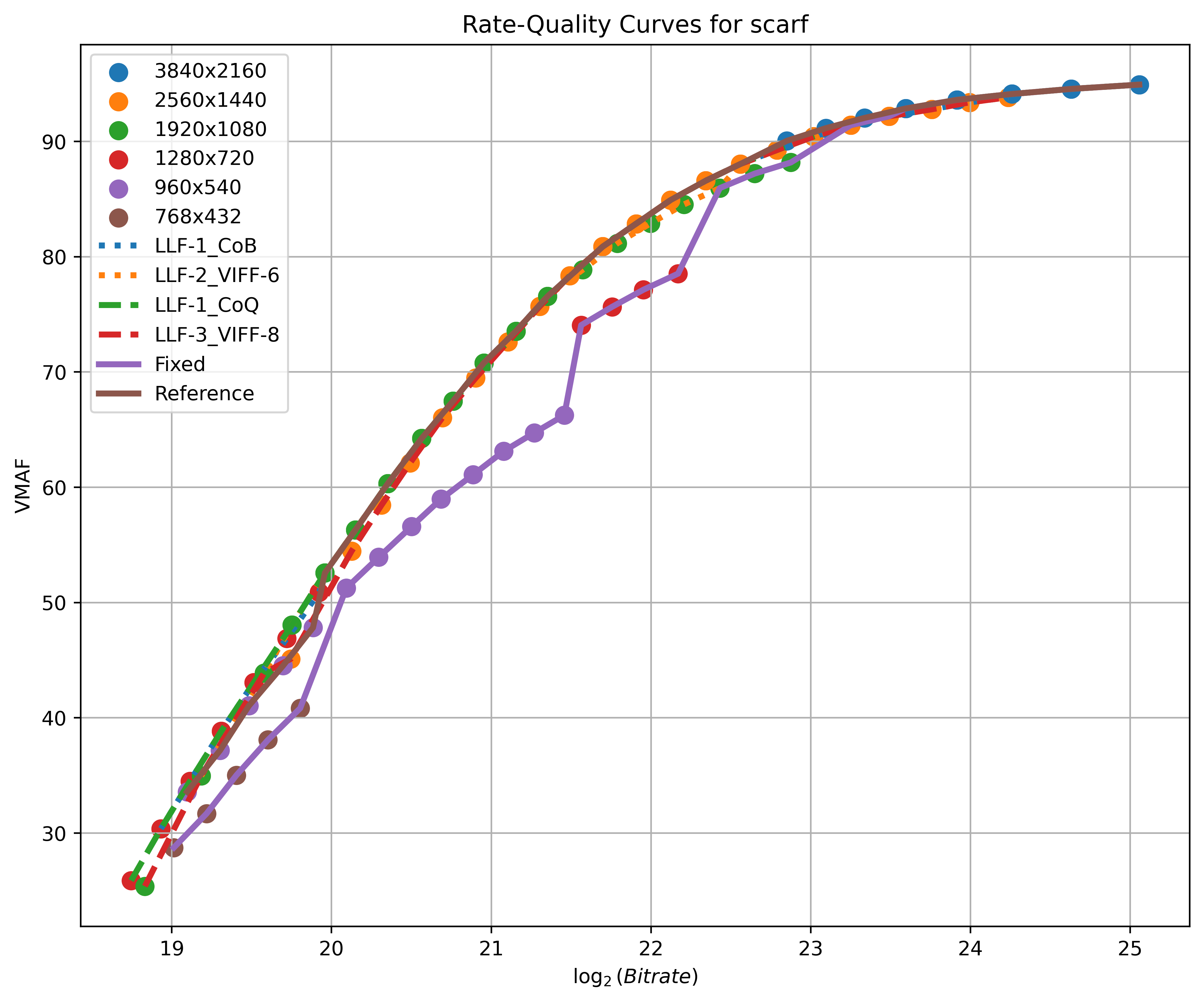}
	\end{subfigure}
	\begin{subfigure}[b]{0.245\linewidth}
	\centering
	\includegraphics[width=\linewidth]{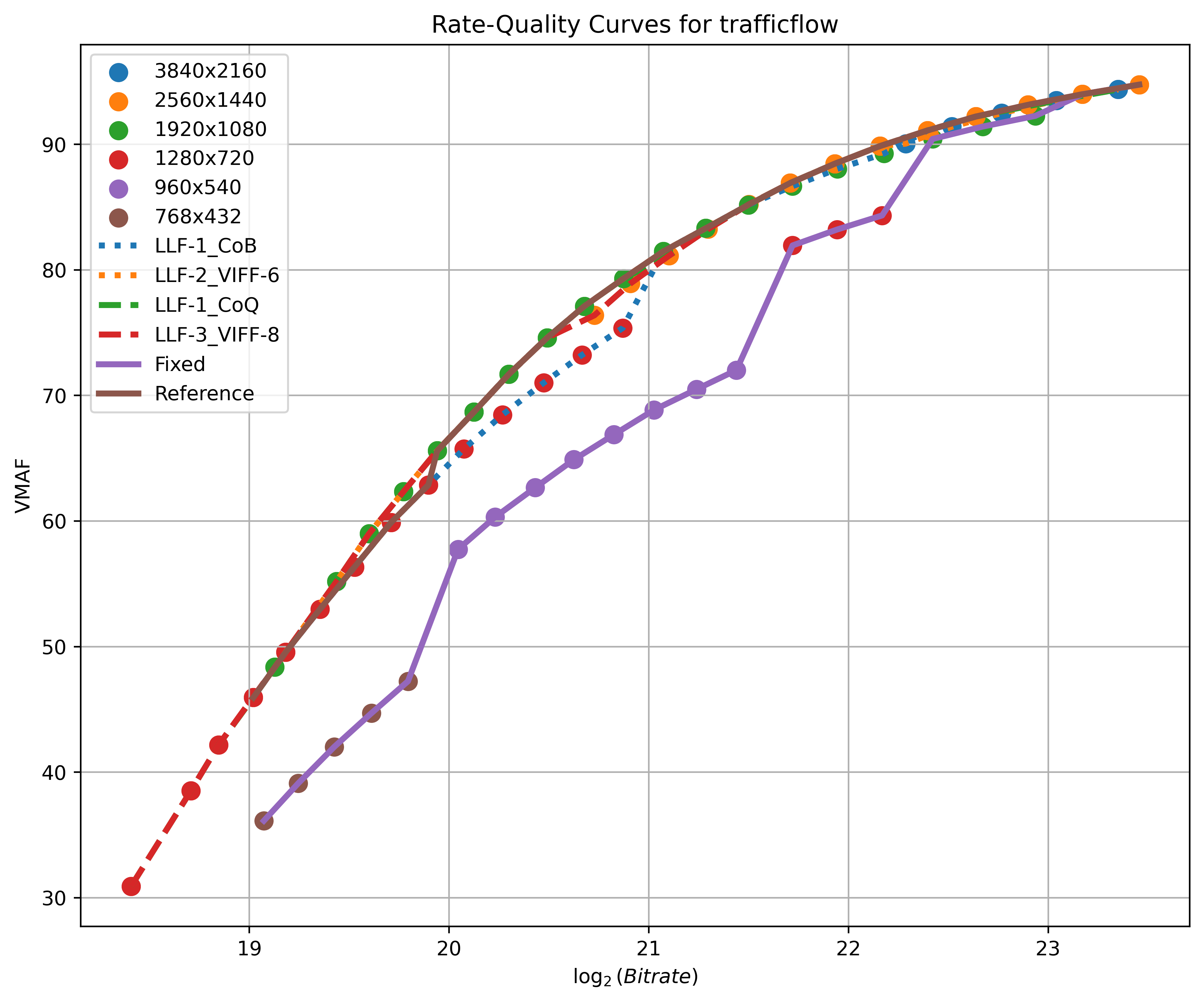}
	\end{subfigure}
	\begin{subfigure}[b]{0.245\linewidth}
	\centering
	\includegraphics[width=\linewidth]{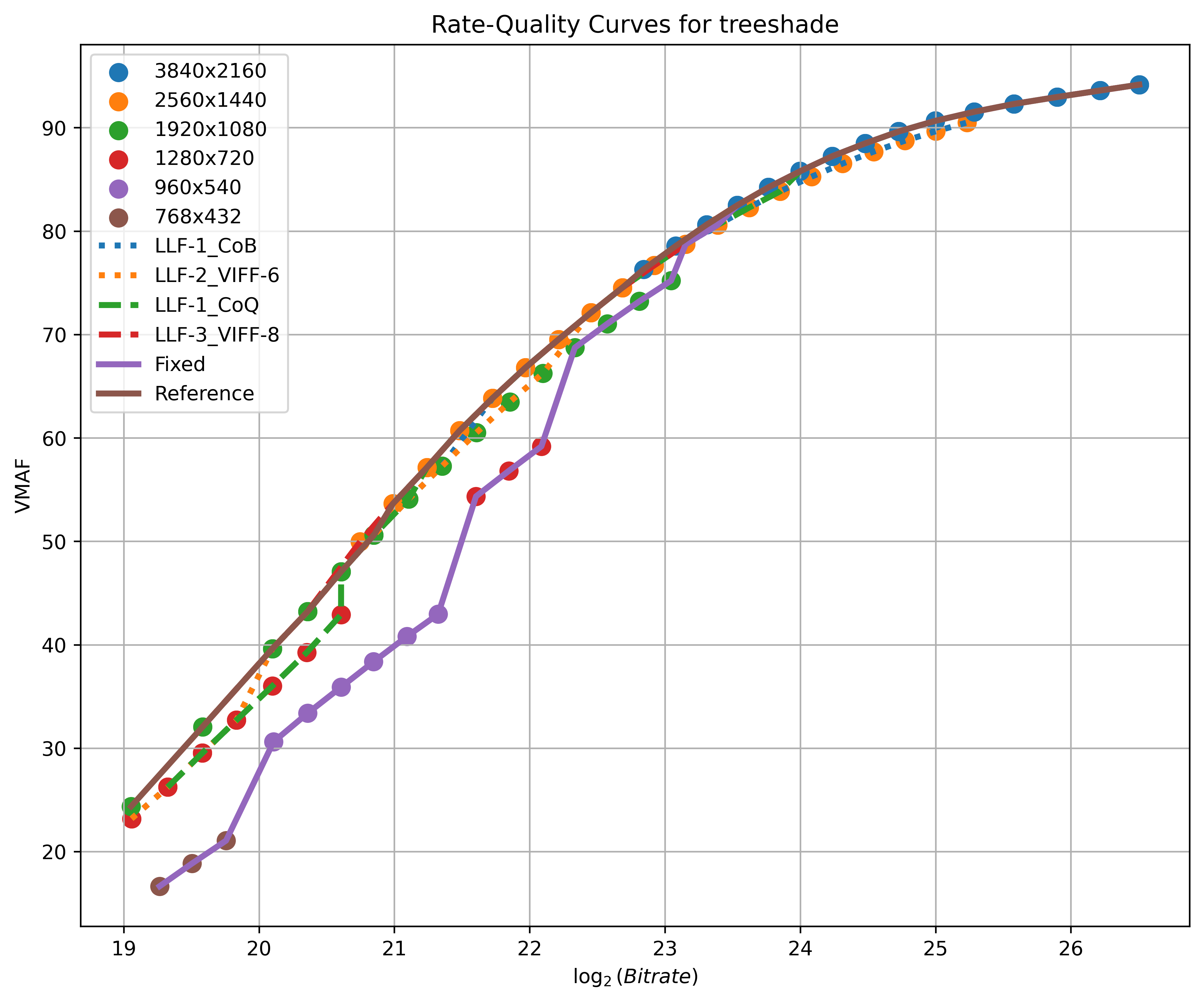}
	\end{subfigure}
	\begin{subfigure}[b]{0.245\linewidth}
	\centering
	\includegraphics[width=\linewidth]{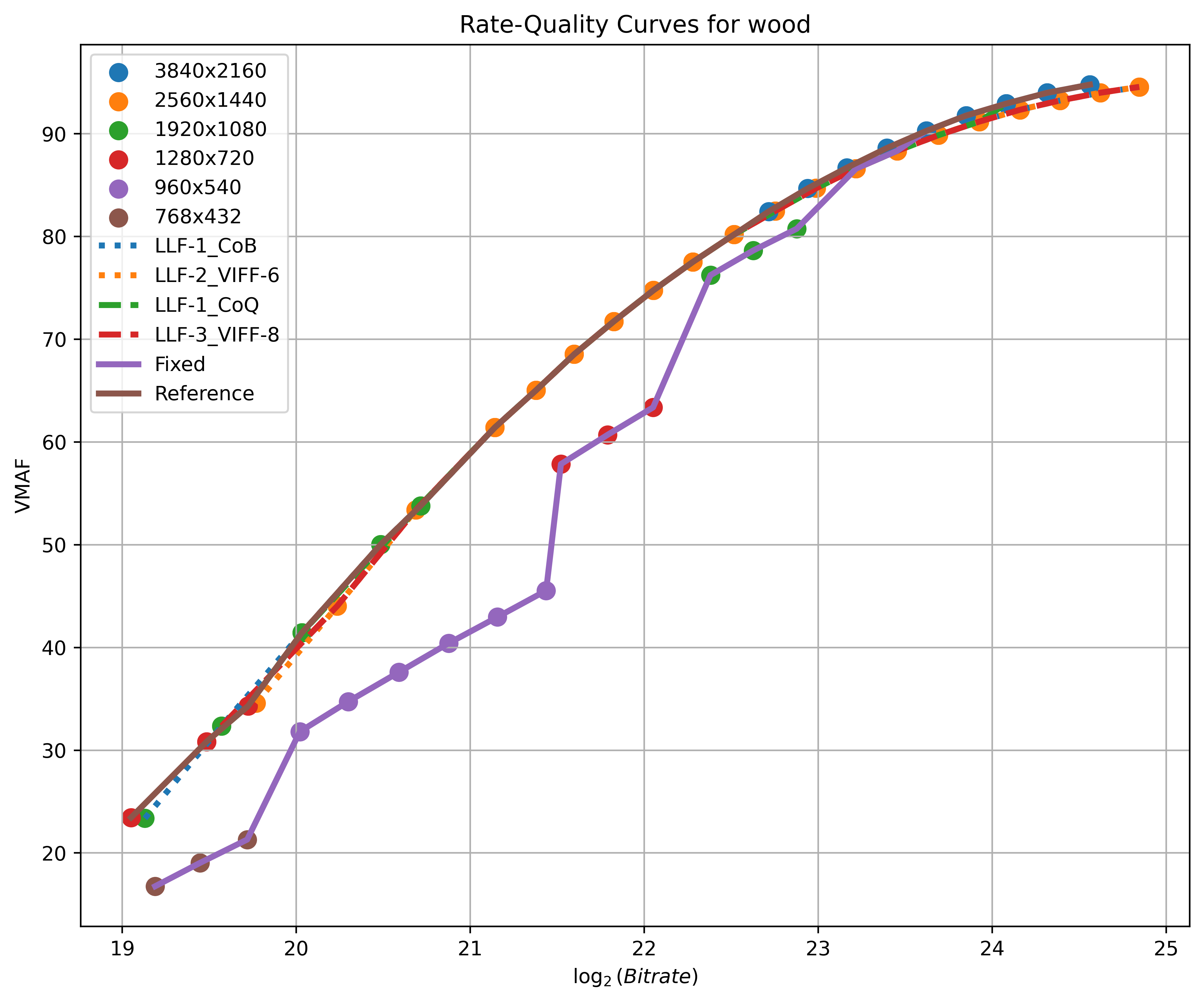}
	\end{subfigure}
	\begin{subfigure}[b]{0.245\linewidth}
	\centering
	\includegraphics[width=\linewidth]{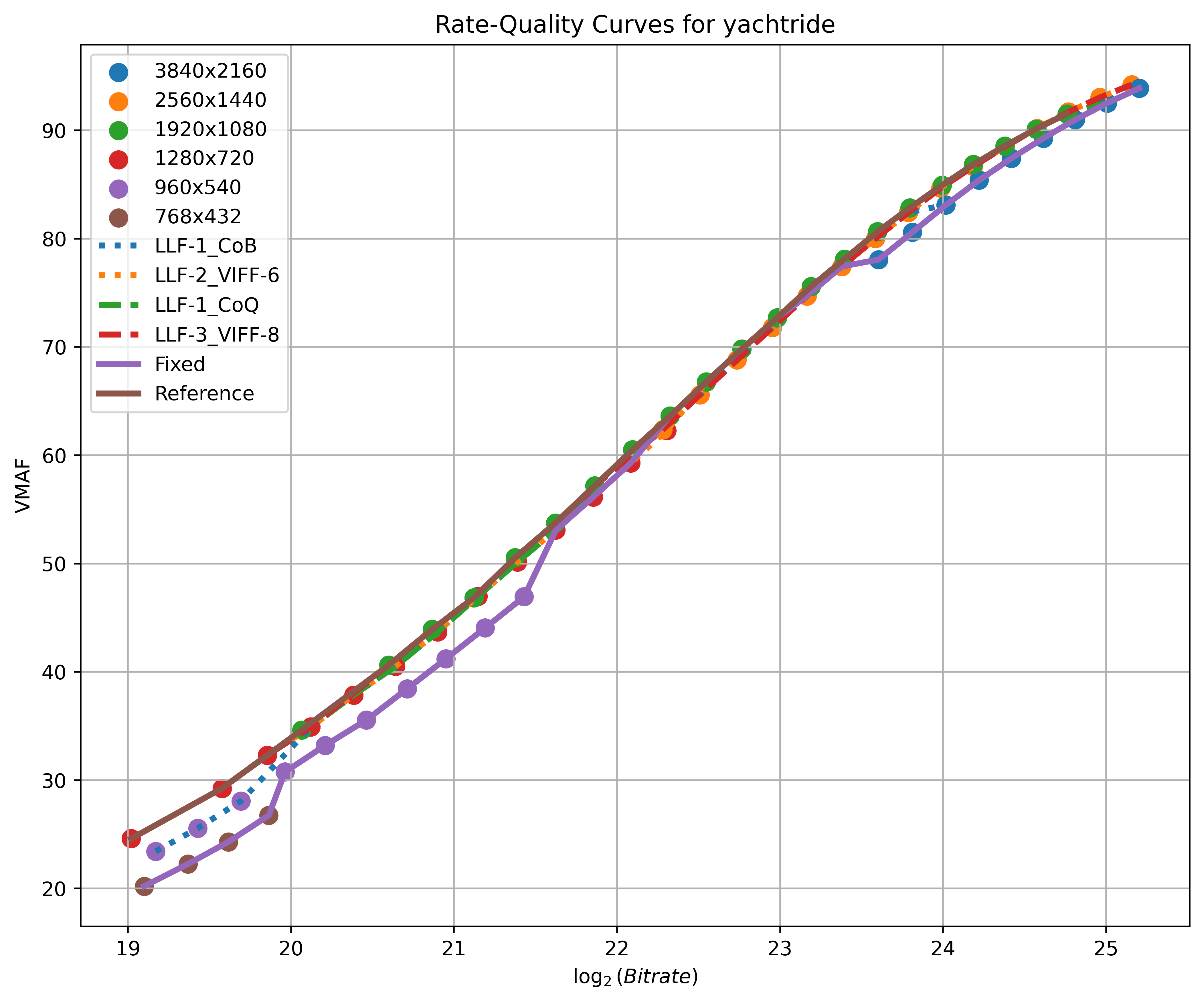}
	\end{subfigure}
	\caption{Convex hulls constructed using ladders from the best performing prediction models, the fixed bitrate ladder, and the reference bitrate ladder on video samples from the validation and test datasets.}
	\label{fig:predicted_rate_quality_curves}
\end{figure*}

One of the salient differences between per-shot bitrate and quality ladders constructed by predicting quality and bitrate (respectively) is that predicted per-shot bitrate ladders struggle at low bitrates and include fewer low resolutions, while the predicted per-shot quality ladders tend to struggle at higher qualities and include fewer high resolutions. This can be ascribed to errors in the ML models during prediction. For this reason, the BD-Rate gains demonstrated by the quality ladders were slightly higher than the BD-Rate gains of bitrate ladders, against a fixed bitrate ladder, since the quality ladders contain fewer high resolutions. The right selection of training and test videos, coupled with careful consideration of experimental settings, these techniques hold promise for delivering remarkable performance. Fig. \ref{fig:predicted_rate_quality_curves} shows examples of convex hulls constructed using best-performing bitrate ladders, quality ladders, and fixed and reference bitrate ladders.

\subsection{Complexity}
\begin{table}
	\normalfont
	\normalsfcodes
	\renewcommand{\arraystretch}{1.75}
	\centering
	\caption{Approximate execution times of the various processes involved in the construction of the per-shot ladders using various methods.}
	\label{table:execution-times}
	\resizebox{\columnwidth}{!}{
	\begin{tabular}{| m{12em} | m{8em} | m{8em} | m{8em} |}
		\hline
		\textbf{Process} & \textbf{Language} & \textbf{Mean Execution Time} (in minutes) & \textbf{Max Execution Time} (in minutes) \\
		\hline
		Downsampling and Compression & C & 0.42 & 5.25 \\
		\hline
		Upsampling and VMAF & C & 0.67 & 0.94 \\
		\hline
		$\text{GLCM}$ & Python & 2.40 & - \\
		\hline
		$\text{TC}$ & Python & 1.90 & - \\
		\hline
		$\text{SI, TI, CF}$ & Python & 1.20 & - \\
		\hline
		$\text{CTI, CI}$ & Python & 0.60 & - \\
		\hline
		$\text{Texture-DCT}$ & Python & 0.80 & - \\
		\hline
		$\text{VIFF}_{3}$ & Python & 20.00 & - \\
		\hline
		$\text{VIFF}_{9}$ & Python & 40.00 & - \\
		\hline
		$\text{LLF}_{1}$ & Python & 6.00 & - \\
		\hline
		Extra-Trees Regressor & Python & 0.10 & - \\
		\hline
	\end{tabular}}
\end{table}

A significant challenge associated with the utilization of the Dynamic Optimizer \cite{Dynamic-Optimizer, DO} is its substantial computational demand. Construction of a convex hull for an individual shot requires compressing the shot at many resolutions and rate control settings. Subsequently, each compressed shot must be upsampled to its original resolution to compute VMAF. This procedure imposes a considerable time and resource burden. As discussed earlier, a variety of techniques have been proposed to optimize this procedure. In our experiments, we utilized an AMD Ryzen 9 5950X and reported the total execution time, which includes both user and system time. The execution time for compression, as well as the combined execution time for upsampling and quality estimation, varied depending on the resolution and the Constant Rate Factor (CRF) used during video encoding. As a result, we present the mean and maximum observed execution times.

Table \ref{table:execution-times} shows the mean and maximum execution times of various procedures involved in the construction of per-shot bitrate/quality ladders using our proposed method, versus constructing reference bitrate ladders using exhaustive encoding. When utilizing \textit{ffmpeg} with the libx265 codec and a medium preset, we observed an approximate mean execution time of 0.42 minutes and a maximum execution time of 5.25 minutes for compression. We observed a mean combined execution time of approximately 0.67 minutes, with a maximum of around 0.94 minutes, for the upsampling and quality estimation processes. Consequently, the construction of a reference bitrate ladder sampled from a convex hull via exhaustive encoding would require approximately 150 minutes on average, given our experimental settings. We also observed that the execution time to compute the low-level feature set $\text{LLF}_{1}$ required about 6 minutes, while calculating the VIF features $\text{VIFF}_{9}$ needed about 40 minutes. The execution time consumed by the Extra-Trees \cite{Extra-Trees} regressor was approximately 0.1 minutes per video. Table \ref{table:execution-times} also shows the execution times to compute various subsets of low-level features and VIF features.

Of course, the execution of Python-based code is considerably slower than that of C-based code (\textit{ffmpeg}). This disparity becomes particularly noticeable when comparing the execution time of VMAF to the calculation of VIF features. Estimating VMAF between reference and compressed videos involves calculating VIF features on both videos. By contrast, VIF features in our models are only calculated on the reference video, which is the uncompressed video. When \textit{ffmpeg} is used, due to its C-based execution, this process requires a maximum of approximately 0.94 minutes. By comparison, calculating VIF features in Python on a video sample required around 40 minutes. It is also worth noting that these execution times varied depending on video complexity, and the total execution time (user + system) was different from the time elapsed. Therefore, the execution times discussed should be understood as providing a baseline understanding of the time complexities of each process, but do not necessarily represent the best achievable execution time.

\begin{table}
	\centering
	\caption{Pearson correlation coefficients between true VMAF and VMAF predicted using various subsets of low-level features at each resolution of the validation and test datasets.}
	\renewcommand{\arraystretch}{1.75}
	\resizebox{\columnwidth}{!}{
	\begin{tabular}{| c | c | c | c | c | c | c |} 
	\hline
	\textbf{Features} & \textbf{2160p} & \textbf{1440p} & \textbf{1080p} & \textbf{720p} & \textbf{540p} & \textbf{432p}\\ 
	\hline
	\textbf{GLCM, TC} & 0.529 & 0.595 & 0.631 & 0.646 & 0.627 & 0.588\\ 
	\hline
	\textbf{SI, TI} & 0.540 & 0.606 & 0.634 & 0.646 & 0.618 & 0.593\\ 
	\hline
	\textbf{CTI, CF, CI} & 0.440 & 0.494 & 0.527 & 0.551 & 0.500 & 0.385\\ 
	\hline
	\textbf{DCT-Texture, Bitrate-DCT-Texture} & 0.619 & 0.673 & 0.690 & 0.674 & 0.616 & 0.522\\
	\hline
	\textbf{$\text{LLF}_{2}$} & 0.580 & 0.642 & 0.668 & 0.677 & 0.643 & 0.602\\
	\hline
	\end{tabular}}
	\label{table:LLF-Ablation-Results-Quality-Prediction}
\end{table}

\begin{table}
	\centering
	\caption{Pearson correlation coefficients between true bitrate and bitrate predicted using various subsets of low-level features at each resolution of the validation and test datasets.}
	\renewcommand{\arraystretch}{1.75}
	\resizebox{\columnwidth}{!}{
	\begin{tabular}{| c | c | c | c | c | c | c |} 
	\hline
	\textbf{Features} & \textbf{2160p} & \textbf{1440p} & \textbf{1080p} & \textbf{720p} & \textbf{540p} & \textbf{432p}\\ 
	\hline
	\textbf{GLCM, TC} & 0.538 & 0.569 & 0.581 & 0.582 & 0.554 & 0.517\\ 
	\hline
	\textbf{SI, TI} & 0.615 & 0.647 & 0.663 & 0.678 & 0.671 & 0.655\\ 
	\hline
	\textbf{CTI, CF, CI} & 0.500 & 0.525 & 0.536 & 0.520 & 0.463 & 0.394\\ 
	\hline
	\textbf{DCT-Texture, VMAF-DCT-Texture} & 0.601 & 0.652 & 0.684 & 0.703 & 0.701 & 0.688\\
	\hline
	\textbf{$\text{LLF}_{3}$} & 0.565 & 0.598 & 0.617 & 0.626 & 0.613 & 0.595\\
	\hline
	\end{tabular}}
	\label{table:LLF-Ablation-Results-Bitrate-Prediction}
\end{table}

\begin{table*}
	\normalfont
	\normalsfcodes
	\renewcommand{\arraystretch}{1.75}
	\centering
	\caption{Means and standard deviations of BD-metrics, and closeness of each model's predicted per-shot bitrate ladders against fixed and reference bitrate ladders on the validation and test datasets, trained on various subsets of low-level features.}
	\resizebox{\textwidth}{!}{
	\begin{tabular}{| m{4.25cm} | c | c | c | c | c | c | c |}

	\hline
	\textbf{Features Set} & \multicolumn{2}{ c |}{\textbf{BL vs Fixed Bitrate Ladder}} & \multicolumn{2}{ c |}{\textbf{BL vs Reference Bitrate Ladder}} & $\textbf{f}_{25}$ & $\textbf{f}_{50}$ & $\textbf{f}_{75}$ \\
	\cline{2-5}
	& BD-Rate (in \%) & BD-VMAF & BD-Rate (in \%) & BD-VMAF & & &\\
	\hline

	GLCM, TC, $\log_{2}(b)$, $\frac{w}{3840}$, $\frac{h}{3840}$ & $-19.860/17.132$ & $4.116/3.887$ & $1.509/4.051$ & $-0.203/0.705$ & $0.933$ & $0.833$ & $0.767$\\

	\hline

	SI, TI, $\log_{2}(b)$, $\frac{w}{3840}$, $\frac{h}{3840}$ & $-19.044/16.788$ & $4.136/4.090$ & $2.046/5.831$ & $-0.248/0.987$ & $0.900$ & $0.800$ & $0.733$\\

	\hline

	CTI, CF, CI, $\log_{2}(b)$, $\frac{w}{3840}$, $\frac{h}{3840}$ & $-17.498/21.019$ & $3.736/4.789$ & $2.824/6.079$ & $-0.596/1.514$ & $0.867$ & $0.800$ & $0.733$\\

	\hline

	DCT-Texture, Bitrate-DCT-Texture, $\log_{2}(b)$, $\frac{w}{3840}$ & $-19.464/18.260$ & $4.066/4.210$ & $1.197/4.246$ & $-0.191/0.815$ & $0.867$ & $0.800$ & $0.767$\\

	\hline

	$\text{LLF}_{2}$, $\log_{2}(b)$, $\frac{w}{3840}$, $\frac{h}{3840}$ & $-20.429/16.747$ & $4.323/4.092$ & $0.439/4.791$ & $-0.016/1.004$ & $0.967$ & $0.933$ & $0.767$\\

	\hline
	\end{tabular}}
	\label{table:bitrate-ladder-bd-metrics-LLF}
\end{table*}

\begin{table*}
	\normalfont
	\normalsfcodes
	\renewcommand{\arraystretch}{1.75}
	\centering
	\caption{Means and standard deviations of BD-metrics, and closeness of each model's predicted per-shot quality ladders against fixed and reference bitrate ladders on the validation and test datasets, trained on various subsets of low-level features.}
	\resizebox{\textwidth}{!}{
	\begin{tabular}{| m{4.25cm} | c | c | c | c | c | c | c |}

	\hline
	\textbf{Features Set} & \multicolumn{2}{ c |}{\textbf{QL vs Fixed Bitrate Ladder}} & \multicolumn{2}{ c |}{\textbf{QL vs Reference Bitrate Ladder}} & $\textbf{f}_{25}$ & $\textbf{f}_{50}$ & $\textbf{f}_{75}$ \\
	\cline{2-5}
	& BD-Rate (in \%) & BD-VMAF & BD-Rate (in \%) & BD-VMAF & & & \\
	\hline

	GLCM, TC, $q/100$, $\frac{w}{3840}$, $\frac{h}{3840}$ & $-20.204/16.671$ & $4.462/4.117$ & $0.975/3.653$ & $-0.118/0.776$ & $0.967$ & $0.933$ & $0.833$\\

	\hline

	SI, TI, $q/100$, $\frac{w}{3840}$, $\frac{h}{3840}$ & $-19.595/16.347$ & $4.300/4.062$ & $1.487/3.430$ & $-0.191/0.741$ & $0.967$ & $0.900$ & $0.800$\\

	\hline

	CTI, CF, CI, $q/100$, $\frac{w}{3840}$, $\frac{h}{3840}$ & $-18.816/16.868$ & $4.105/4.156$ & $2.545/5.125$ & $-0.368/1.095$ & $0.867$ & $0.800$ & $0.700$\\

	\hline

	DCT-Texture, VMAF-DCT-Texture, $q/100$, $\frac{w}{3840}$, $\frac{h}{3840}$ & $-19.904/16.712$ & $4.454/4.072$ & $1.184/4.086$ & $-0.152/0.877$ & $0.933$ & $0.833$ & $0.767$\\

	\hline

	$\text{LLF}_{3}$, $q/100$, $\frac{w}{3840}$, $\frac{h}{3840}$ & $-20.546/16.722$ & $4.599/4.107$ & $0.615/3.546$ & $-0.05/0.786$ & $1.000$ & $0.900$ & $0.900$\\

	\hline
	\end{tabular}}
	\label{table:quality-ladder-bd-metrics-LLF}
\end{table*}

\section{Conclusion and Future Work}
\label{sec:conclusion}
We studied multiple feature sets based on VIF features and an ensemble of VIF features and low-level features, for constructing content-gnostic ladders. Our proposed methods predict per-shot bitrate or quality ladders of a video using regressors trained to predict video quality or bitrate of compressed videos using video features and metadata, respectively. We compared our proposed models against existing popular techniques, including predicting cross-over bitrates and quality using low-level features. Our observations show that regressors trained on low-level features and VIF features demonstrated significant gains in bitrate and quality against the fixed bitrate ladder and small losses against the reference bitrate ladder. Their performance is followed closely by the cross-over point prediction models, regressors trained on low-level features and VIF features, respectively. Our methods predict approximate bitrates and video quality without compression or quality estimation, allowing us to construct bitrate and quality ladders, without relying on exhaustive encoding. These results suggest a significant potential for efficiently constructing per-shot bitrate and quality ladders.

\subsubsection*{Future Work}
In the future, we aim to conduct experiments on larger datasets to better understand and improve per-shot ladder construction techniques. We intend to evaluate the efficacy of these methods across diverse presets and codecs. Furthermore, we will investigate the transferability of constructed per-shot bitrate ladders across multiple encoder settings.

\section*{Acknowledgments}
The authors thank the Texas Advanced Computing Center (TACC) at The University of Texas at Austin for providing HPC resources that have contributed to the research results reported in this paper. URL: http://www.tacc.utexas.edu.

\bibliographystyle{IEEEtran}
\bibliography{refs}

\end{document}